\documentclass[twocolumn,showpacs,preprintnumbers,amsmath,amssymb,aps,pra]{revtex4-1}

\usepackage{graphicx}
\usepackage{dcolumn} 
\usepackage{bm, bbm}      
\usepackage{curves}
\usepackage{epic}
\usepackage{wasysym}
\usepackage{sidecap}
\usepackage{multirow}
\usepackage{amssymb, amsfonts, latexsym}
\usepackage[mathcal]{euscript}
\usepackage{epsf, times}
\usepackage{subfigure}
\usepackage{amsthm}

\def\openone{\leavevmode\hbox{\small1\kern-4.2pt\normalsize1}}

\def\S{\mathcal{S}}
\def\A{\mathcal{A}}
\def\B{\mathcal{B}}
\def\Ai{\mathcal{A}_1}
\def\Aii{\mathcal{A}_2}
\def\Bi{\mathcal{B}_1}
\def\Bii{\mathcal{B}_2}

\def\N{\mathcal{N}}
\def\E{\mathcal{E}}

\def\Id{\mathbb{I}}
\def\og{\overline{g}}
\def\oog{\overline{\overline{g}}}
\def\tg{\tilde{g}}

\newcommand{\beq}{\begin{equation}}
\newcommand{\eeq}{\end{equation}}
\newcommand{\bea}{\begin{eqnarray}}
\newcommand{\eea}{\end{eqnarray}}
\newcommand{\bfig}{\begin{figure}}
\newcommand{\efig}{\end{figure}}

\begin{document}

\title{
Distilling topological entropy from single entanglement measures on projected systems
      }

\author{
C. Castelnovo$^{1}$
}
\affiliation{
$^1$
TCM group, 
Cavendish Laboratory, 
University of Cambridge, 
Cambridge CB3 0HE, United Kingdom
}

\date{\today}

\begin{abstract}

Entanglement measures find frequent application in the study 
of topologically ordered systems, where the presence of 
topological order is reflected in an additional contribution to 
the entanglement of the system. 
%
Obtaining this topological entropy from analytical 
calculations or numerical simulations is generally difficult 
due to the fact that it is an order one correction to leading terms 
that scale with the size of the system. 
In order to distil the topological entropy, one resorts 
to extrapolation as a function of system size, or to clever 
subtraction schemes that allow to cancel out the leading terms. 
Both approaches have the disadvantage of requiring multiple 
(accurate) calculations of the entanglement of the system. 
Here we propose a modification of conventional entanglement calculations 
that allows to obtain the topological entropy of a system from a single 
measurement of entanglement. 
In our approach, we replace the conventional trace 
over the degrees of freedom of a partition of the system with a projection 
onto a given state (which needs not be known). 
We show that a proper choice of partition and projective measurement 
allows to rid the entanglement measures of the 
typical boundary terms, thus exposing the topological contribution alone. 
We consider specifically the measures known as 
von Neumann entropy and entanglement negativity, and we discuss their 
application to both models that exhibit quantum as well as classical 
topological order. 

\end{abstract}

\maketitle
%
%

\section{
Introduction
        }
Entanglement measures, such as the von Neumann entropy or the entanglement 
negativity, are often used to study and 
characterise topologically ordered systems. It was indeed 
demonstrated that the presence of topological order gives rise to an 
additional contribution to the entanglement of the system, 
which relates directly to the quantum dimension of its anyonic 
excitations~\cite{Levin2006,Kitaev2006}. 
This contribution was dubbed \textit{topological entanglement entropy}. 

Obtaining the topological entropy of a system from analytical 
calculations or numerical simulations is in general a tall order. 
This is due to the fact that the desired contribution is 
an order one correction to leading terms that scale with the size of the 
system. 
For example, the von Neumann entropy of a bipartition $\S = \A \cup \B$ 
is known to exhibit an \textit{area law} contribution that scales 
with the size of the boundary between $\A$ and $\B$. The topological 
entanglement entropy $\gamma$ is an order one correction to it. 
It was recently demonstrated that the entanglement 
negativity~\cite{Zyczkowski1998,Lee2000,Eisert2001,Plenio2005,Vidal2002} 
behaves in a similar way~\cite{Lee2013,Castelnovo2013}. 

In order to distil the topological entropy, one has to resort 
either to extrapolations as a function of system size 
(see e.g., Ref.~\onlinecite{Stephan2012}) or to clever 
subtraction schemes that allow to cancel out the leading terms and to expose 
the order one topological correction~\cite{Levin2006,Kitaev2006}. 
Both approaches have the disadvantage of requiring multiple 
(and $\mathcal{O}(1)$ accurate) calculations of the entanglement of the 
system. 

Here we propose a modification of conventional entanglement calculations 
that allows us to obtain the topological entropy of a system from a single 
measurement of entanglement. 
The key feature in our approach is to replace the conventional trace 
over the degrees of freedom of a partition of the system with a projection 
onto a given state (which needs not be known), 
\beq
\rho_\A \propto \langle \phi_\B \vert \rho \vert \phi_\B \rangle 
\qquad \textrm{vs.} \qquad 
\rho_\A \propto {\rm Tr}_{\B} \rho 
. 
\nonumber 
\eeq
We show that a proper choice of partitions and projective measurement 
allows to rid the entanglement measures of the 
typical boundary terms, thus exposing the topological contribution alone. 

We consider specifically the measures known as 
von Neumann entropy and entanglement negativity, and we discuss their 
application to models that exhibit quantum as well as classical 
topological order. 
The models of choice will be Kitaev's toric code model and the eight 
vertex model, for they allow an exact calculation of both the von Neumann 
entropy and the entanglement negativity. 

If the von Neumann entanglement entropy is used, the resulting topological 
contribution can be either due to classical or quantum topological 
correlations. For instance, the result is the same for the classical 
eight-vertex model as for the quantum toric code. 
On the other hand, the entanglement negativity is sensitive only to quantum 
topological correlations and gives a non-vanishing result only in the case 
of quantum topological order, as in Kitaev's toric code. 
Therefore, the combined use of von Neumann and negativity calculations 
allows a straighforward detection of topological order in both classical 
and quantum systems, as well as a clear distinction between the two cases. 
%
%

\section{
Toric code and measures of entanglement
        }
The toric code is a system of spin-1/2 degrees of freedom 
$\sigma_i$ living on the bonds $i$ of a square lattice (periodic boundary 
conditions are assumed throughout). 
The Hamiltonian of the system can be written as~\cite{Kitaev2003}: 
\bea
H = - \lambda_A \sum_s A_s - \lambda_B \sum_p B_p 
\label{eq: TC Hamiltonian}
\\
A_s = \prod_{i \in s} \sigma^x_i 
\qquad 
B_p = \prod_{i \in p} \sigma^z_i
, 
\nonumber
\eea
where $s$ ($p$) label the sites (plaquettes) of the lattice. 

The ground state (GS) is 4-fold degenerate, according to the 4 topological 
sectors identified by the expectation values of winding loop operators. 
Within each sector, the GS is given by the equal amplitude superposition of 
all tensor product basis states $\otimes_i \vert \sigma_i^z \rangle$ 
belonging to that sector. 
Following the notation in Refs.~\onlinecite{Hamma2005,Castelnovo2008}, 
we introduce the group $G$ generated by products of $A_s$ operators. 
Notice that one has to define elements $g \in G$ modulo the identity 
$\prod_s A_s = \Id$ in order for the inverse of $g$ to be uniquely defined 
(in which case, $g^{-1} = g$). The order (i.e., the number of elements) 
of $G$ is therefore $\vert G \vert = 2^{N^{(s)}-1}$, where $N^{(s)}$ is 
the number of sites on the lattice. 
If we define $\vert 0 \rangle \equiv \otimes_i \vert \sigma_i^z = +1 \rangle$, 
one of the 4 topologically ordered GS wavefunctions can be written explicitly 
as: 
\beq
\vert \psi_0 \rangle 
= 
\frac{1}{\vert G \vert^{1/2}} 
  \sum_{g \in G} g \vert 0 \rangle 
. 
\label{eq: TC GS}
\eeq
Note that the choice of reference state $\vert 0 \rangle$ is immaterial and 
one can replace $\vert 0 \rangle$ with $\overline{g} \vert 0 \rangle$, 
for any given $\overline{g} \in G$, and the state $\vert \psi_0 \rangle$ 
remains unchanged. 
The other 3 GS wavefunctions are obtained upon choosing reference states 
$\vert 0 \rangle$ that are in different topological sectors with respect 
to $\otimes_i \vert \sigma_i^z = +1 \rangle$. 

The von Neumann entropy $S_{\rm vN}^{(\A)}$ obtained for a bipartition of 
the system $\S = \A \cup \B$ is defined as 
$S_{\rm vN}^{(\A)} = - {\rm Tr} \rho_\A \ln \rho_\A$, 
where $\rho_\A = {\rm Tr}_\B \rho$. 

The negativity $\N$ (or, equivalently, the \emph{logarithmic negativity} 
$\E$), is defined from the trace norm $\| \rho^{T_\B} \|_1$ of the 
partial transpose over subsystem $\B$ of the density matrix $\rho$, 
\bea
\N \equiv \frac{\| \rho^{T_\B} \|_1 - 1}{2} 
, 
\qquad\quad 
\E \equiv \ln \| \rho^{T_\B} \|_1
, 
\label{eq: log negativity}
\eea
where $\| \rho^{T_\B} \|_1$ is the sum of the absolute 
values of the eigenvalues $\lambda_i$ of $\rho^{T_\B}$. 
If all the eigenvalues are positive then $\N = 0$ (recall that 
$\sum_i \lambda_i = 1$) and $\N > 0$ otherwise. 

In the following we use the conventional replica trick to calculate the 
von Neumann entropy (see e.g., 
Refs.~\onlinecite{Hamma2005,Castelnovo2007,Castelnovo2008}). 
We also use a replica approach to compute the negativity of the system 
which was recently introduced by Calabrese, Cardy and 
Tonni~\cite{Calabrese2012} (see also 
Refs.~\onlinecite{Calabrese2013,Alba2013}). 
This replica approach has already been 
applied to the toric code model in Refs.~\onlinecite{Lee2013,Castelnovo2013}. 
%
%

\section{
Partition and projection
        }
\begin{figure}
\includegraphics[width=0.75\columnwidth]{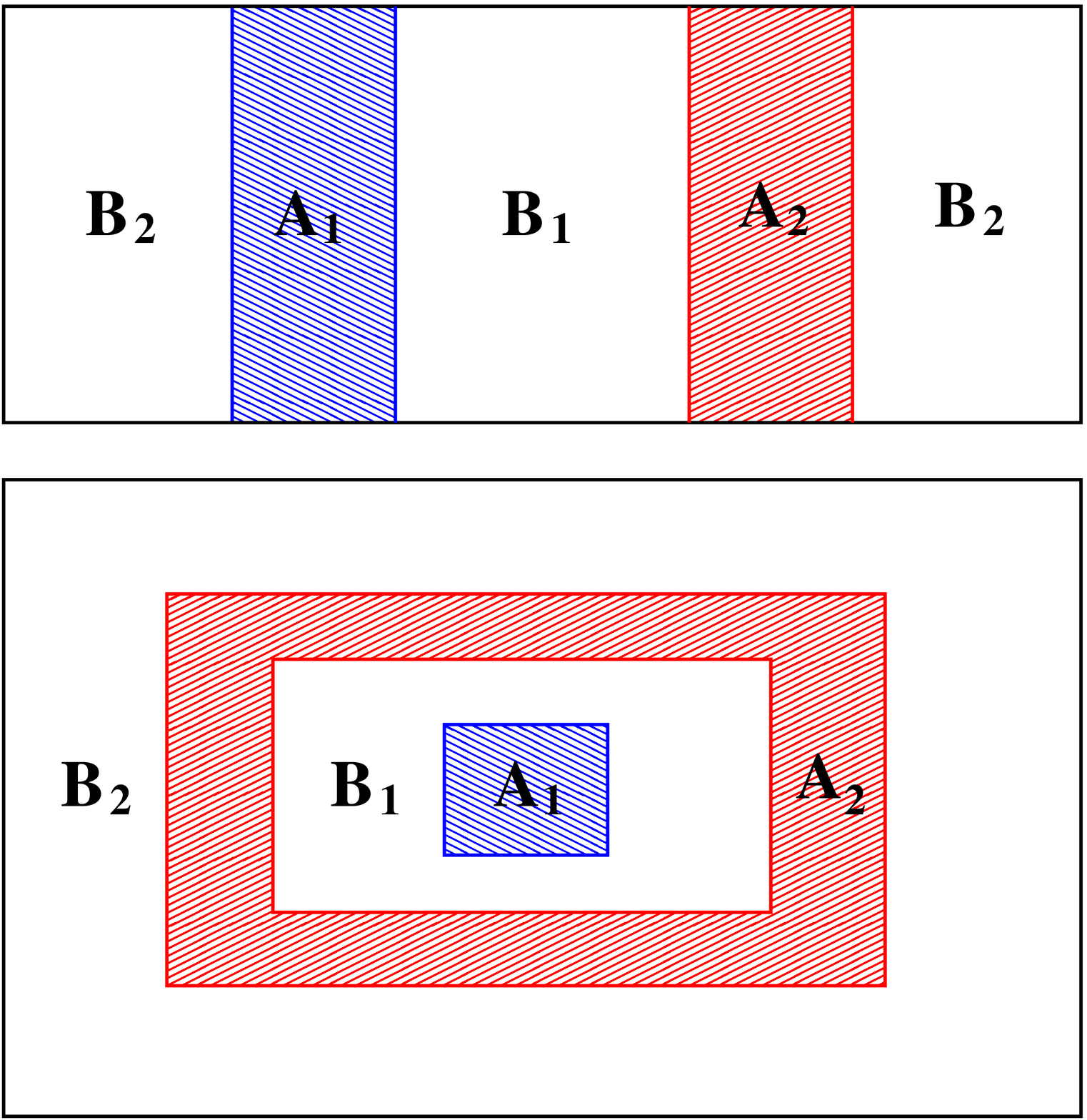}
\caption{
\label{fig: partitions} 
(Colour online) -- 
Examples of partitions of the system $\S = \A \cup \B$, 
where subsystem $\B$ splits $\A$ into two components, $\A = \Ai \cup \Aii$, 
and vice versa. 
}
\end{figure}

The method proposed in this paper requires the use of bi-partitions of a 
system $\S = \A \cup \B$, where subsystem $\B$ splits $\A$ into two 
disconnected components, $\A = \Ai \cup \Aii$, and vice versa, as illustrated 
in Fig.~\ref{fig: partitions}

Without loss of generality, we will assume that the toric code has been 
prepared in the GS $\vert \psi_0 \rangle$ in Eq.~\eqref{eq: TC GS}. 
Earlier results for the von Neumann entropy~\cite{Hamma2005,Castelnovo2008} 
can be straightforwardly applied to show that 
\bea
S_{\rm vN}^{(\A)} 
&=& - {\rm Tr} \rho_\A \ln \rho_\A 
= - \ln \frac{\vert G_\A \vert\,\vert G_\B \vert}{\vert G \vert} 
\\ 
&=& 
\ln \left[ 
  2^{N^{(s)}_{\partial_{\A}} + 1 
	- 
	(m_\A + m_\B)}
\right]
\\ 
S_{\rm vN}^{(\Ai)} 
&=& - {\rm Tr} \rho_{\Ai} \ln \rho_{\Ai} 
= - \ln \frac{\vert G_{\Ai} \vert\,\vert G_{\Aii \cup \B} \vert}{\vert G \vert}
\\
&=& 
\ln \left[ 
  2^{N^{(s)}_{\partial_{\Ai}} - 1 }
\right]
, 
\eea
where $G_\alpha$ is the subgroup of $G$ that acts only on spins in subsystem 
$\alpha$ 
($G_\alpha = \{ g \in G \, \vert \, g=g_\alpha \otimes \openone_{\alpha^c}\}$, 
$\S = \alpha \cup \alpha^c$), 
and $\vert G_\alpha \vert$ is its order. 
$N^{(s)}_{\partial_{\alpha}} \equiv N^{(s)}_{\partial_{\alpha^c}}$ 
is the number of star operators acting simultaneously on spins in $\alpha$ 
and on spins in its complementary subsystem $\alpha^c$. 
Moreover, $m_{\alpha}$ is the number of disconnected components of subsystem 
$\alpha$. 

In order to arrive at the results above, we used the fact that 
$\vert G_{\Ai} \vert = 2^{N^{(s)}_{\Ai}}$ and 
$
\vert G_{\Aii \cup \B} \vert 
= 
2^{N^{(s)}_{\Bi} + N^{(s)}_{\Bii} + N^{(s)}_{\Aii} + N^{(s)}_{\partial\Aii}}
$ 
(both subsystems are topologically trivial), whereas 
$
\vert G_{\A} \vert 
= 
2^{N^{(s)}_{\Ai} + N^{(s)}_{\Aii} + m_\B - 1}
$
and 
$
\vert G_{\B} \vert 
= 
2^{N^{(s)}_{\Bi} + N^{(s)}_{\Bii} + m_\A - 1} 
$ 
(both subsystems are topologically non-trivial). 
The topological nature of the system is reflected in the appearance of the 
contributions $m_\A = 2$ and $m_\B = 2$, which depend solely on the topology 
of the partition, as we explain hereafter. 

The factor $2^{m_{\alpha^c}-1}$ in $\vert G_{\alpha} \vert$ is due to the 
fact that, if $\alpha^c$ has more than one disconnected component, then 
the product of all star operators in each component $\alpha^c_i$ times 
the product of all star operators of its boundary $\partial_{\alpha^c_i}$ 
is an operation acting solely on $\alpha$ that \textit{cannot} be written 
in terms of star operators in $\alpha$. 
This is perhaps best illustrated by looking at the example in 
Fig.~\ref{fig: partitions}, top panel. 
Consider the product of all star operators acting solely on $\Ai$ times 
the product of all star operators at its boundary, i.e., acting 
simultaneously on spins in $\Ai$ and in $\B$. The resulting operators acts 
twice on each and every spin in $\Ai$, thus leaving it unchanged; it acts 
however on some spins in $\B$, and this action cannot be written in terms 
of products of star operators acting on $\B$ alone. Therefore, it is an 
additional operation in $G_{\B}$ independent from the $2^{N^{(s)}_\B}$ that 
one can straightforwardly construct from star operators acting only on $\B$. 
This happens because $m_{\A} > 1$: if $m_{\A} = 1$ then the operation 
described above is in fact equivalent to the product of \textit{all} star 
operators acting only on $\B$, hence the subtraction ``$-1$'' in the exponent 
of the factor $2^{m_{\A}-1}$ appearing in $\vert G_{\B} \vert$. 

Whichever the choice to calculate the von Neumann entropy, the topological 
contribution (of order one) is subordinate to a boundary term 
$N^{(s)}_{\partial_\alpha}$ that scales with the size of the partition. 

Similar results were obtained recently for the entanglement 
negativity~\cite{Lee2013,Castelnovo2013}. If one considers 
$\rho_\A = \textrm{Tr}_\B \rho$ and then computes the trace norm 
$\| \rho^{T_{\Aii}} \|_1$ (after transposing the degrees of freedom in 
$\Aii$), the choice of partitions in Fig.~\ref{fig: partitions} actually 
leads to $\| \rho^{T_{\Aii}} \|_1 = 1$ and vanishing negativity 
(see Refs.~\onlinecite{Lee2013,Castelnovo2013}, or 
App.~\ref{app: alternative negativity} for an alternative derivation). 
It is only when the partitions $\Ai$ and $\Aii$ share a boundary and $\A$ 
spans the system in both directions that one can see a topological 
contribution in the entanglement negativity. However, this is once again 
an order one correction to a boundary contribution that scales with the 
size of the partition. 

In order to remove the boundary contribution without throwing the topological 
baby with the bath water, we consider here the density matrix of subsystem 
$\A$ after performing a projective measurement on the spins in subsystem $\B$: 
\beq
\rho_\A \propto \langle \phi_\B \vert \rho \vert \phi_\B \rangle 
, 
\eeq
where $\vert \phi_\B \rangle$ is taken to be a generic tensor product 
basis state $\otimes_{i\in\B} \vert \sigma^z_i \rangle$ appearing in the 
GS superposition in Eq.~\eqref{eq: TC GS}. 
Namely, $\vert \phi_\B \rangle = \og_\B \vert 0_\B \rangle$, 
for some $\og \in G$. 
This is equivalent, for instance, to performing a measurement of the 
$\sigma^z$ component of each spin in $\B$. 

The results that follow are independent of the choice of $\og$, and knowledge 
of the corresponding state $\vert \phi_\B \rangle$ is immaterial. 
Therefore, we only need to know that the spins in $\B$ have been measured, 
but we do not need to know the result of that measurement. 

Given that the toric code is symmetric upon exchange of $\sigma^z$ and 
$\sigma^x$ operators, whilst exchanging also stars with plaquettes 
(mapping the direct lattice to the face-centred dual lattice), 
the results hold true also if the 
projective measurement is done on the $\sigma^x$ component of the 
spins in $\B$. 
The results however do not generically apply if $\vert \phi_\B \rangle$ is 
a superposition of tensor product states in the $\sigma^z$ basis. 
A counter example and related discussion of the conditions under which 
a superposition is admissible are given in Sec.~\ref{sec: generic state}. 

With the choice $\vert \phi_\B \rangle = \og_\B \vert 0_\B \rangle$, 
for some $\og \in G$, after a few lines of algebra reported in 
App.~\ref{app: rho projected} for convenience, one arrives at the 
expression 
\bea
\rho_\A = \frac{1}{\vert G_\A \vert} 
\sum_{g,g' \in g_\A}
g_\A \vert 0_\A \rangle\langle 0_\A \vert g'_\A 
\label{eq: rho_A proj}
\eea
where $\rho_\A$ has been normalised so that $\textrm{Tr}(\rho_\A) = 1$. 
The state $\vert 0 \rangle$ in the expression above is determined (in part) 
by the choice of $\vert \phi_\B \rangle$. However, as we shall see in the 
following, this plays no role in the calculation of the von Neumann 
entropy and entanglement negativity. 
%
%

\subsection{
Entanglement entropy
           }
Instead of computing the von Neumann entropy directly from $\rho_\A$ in 
Eq.~\eqref{eq: rho_A proj}, which trivially vanishes 
(see App.~\ref{app: S_vN projected}), 
let us trace out the degrees of freedom in $\Aii$, 
\bea
\rho_{\Ai} 
&=& 
\frac{1}{\vert G_\A \vert} \sum_{g,g' \in G_\A} 
g_{\Ai} \vert 0_{\Ai} \rangle\langle 0_{\Ai} \vert g'_{\Ai} 
\nonumber\\
&\times&
\langle 0_{\Aii} \vert g'_{\Aii} g_{\Aii} \vert 0_{\Aii} \rangle
. 
\eea

Keeping $g$ fixed, we notice that the mapping $g' \to \tg = g' g$ is 
one-to-one in $G_\A$, and therefore we can re-write the summation over 
$g'$ as a summation over $\tg$ upon replacing $g' = \tg g$ (recall that 
$g^2 = \openone$ for all $g \in G$). 
The expectation value in the expression above reduces then to the 
condition that $\tg_{\Aii} = \openone_{\Aii}$, or equivalently 
$\tg \in G_{\Ai}$: 
\bea
\rho_{\Ai} 
&=& 
\frac{1}{\vert G_\A \vert} \sum_{g\in G_\A, \: \tg \in G_{\Ai}} 
g_{\Ai} \vert 0_{\Ai} \rangle\langle 0_{\Ai} \vert g_{\Ai} \tg_{\Ai} 
. 
\label{eq: rho_A1}
\eea

We can then compute 
$
S_{\rm vN}^{(\Ai)} 
= 
- {\rm Tr} [ \rho_{\Ai} \!\ln \rho_{\Ai} ] 
$ 
using the replica trick 
$- \lim_{n \to 1} \partial_n [{\rm Tr} (\rho_{\Ai}^n)]$. 
In order to do so, we start by considering 
\bea
\rho_{\Ai}^2 
&=& 
\frac{1}{\vert G_\A \vert^2} 
\sum_{g,g'\in G_\A} 
\sum_{\tg,\tg' \in G_{\Ai}} 
g_{\Ai} \vert 0_{\Ai} \rangle
\langle 0_{\Ai} \vert g'_{\Ai} \tg'_{\Ai} 
\nonumber \\ 
&\times& 
\langle 0_{\Ai} \vert g_{\Ai} \tg_{\Ai} g'_{\Ai} \vert 0_{\Ai} \rangle
. 
\eea
Once again, given $g$ and $\tg$ we can use a one-to-one mapping to 
replace $g' \to g''= g \tg g'$, with $g'' \in G_\A$. 
The expectation value reduces to 
$\langle 0_{\Ai} \vert g''_{\Ai} \vert 0_{\Ai} \rangle$, 
which imposes $g'' \in G_{\Aii}$: 
\bea
\rho_{\Ai}^2 
&=& 
\frac{1}{\vert G_\A \vert^2} 
\sum_{g\in G_\A , g'' \in G_{\Aii}} 
\sum_{\tg,\tg' \in G_{\Ai}} 
\nonumber \\
&\times& 
g_{\Ai} \vert 0_{\Ai} \rangle
\langle 0_{\Ai} \vert g''_{\Ai} g_{\Ai} \tg_{\Ai} \tg'_{\Ai} 
\nonumber \\ 
&=& 
\frac{\vert G_{\Ai} \vert \, \vert G_{\Aii} \vert}{\vert G_\A \vert^2} 
\sum_{g\in G_\A} 
\sum_{\tg \in G_{\Ai}} 
g_{\Ai} \vert 0_{\Ai} \rangle
\langle 0_{\Ai} \vert g_{\Ai} \tg_{\Ai} 
\nonumber \\
&=& 
\frac{\vert G_{\Ai} \vert \, \vert G_{\Aii} \vert}{\vert G_\A \vert} 
\rho_{\Ai}
. 
\label{eq: rho_A1^2 propto rho_A1}
\eea
Here we used the fact that $g''_{\Ai} = \openone_{\Ai}$; the dependence on 
$g''$ disappears, allowing to sum over it and resulting in the factor 
$\vert G_{\Aii} \vert$. 
Moreover, the product $\tg \tg'$ is a generic element 
of $G_{\Ai}$ and therefore we can trivially sum over, say, $\tg'$, 
resulting in the factor $\vert G_{\Ai} \vert$. 

Iterating this identity we obtain 
\beq
{\rm Tr}\left( \rho_{\Ai}^n \right) = 
\left( 
  \frac{\vert G_{\Ai} \vert \, \vert G_{\Aii} \vert}{\vert G_\A \vert}
\right)^{n-1}
\eeq
and 
$
S_{\rm vN}^{(\Ai)} 
= 
- \ln \left( 
\vert G_{\Ai} \vert \, \vert G_{\Aii} \vert 
/ 
\vert G_\A \vert
\right)
$. 
As discussed above, $\Ai$ and $\Aii$ are topologically trivial and therefore 
$\vert G_{\Ai} \vert = 2^{N^{(s)}_{\Ai}}$ and 
$\vert G_{\Aii} \vert = 2^{N^{(s)}_{\Aii}}$. On the other hand, $\A$ 
divides subsystem $\B$ in two disconnected components, and therefore 
$\vert G_{\A} \vert = 2^{N^{(s)}_{\A}} + m_{\B} - 1$, with $m_{\B} = 2$ 
and $N^{(s)}_{\A} = N^{(s)}_{\Ai} + N^{(s)}_{\Aii}$. 
We finally obtain $S_{\rm vN}^{(\Ai)} = \ln 2$, which is indeed the expected 
value of the topological entropy, \textit{without any boundary contribution}. 
%
%

\subsection{
Negativity
           }
Let us now consider the entanglement negativity between subsystem $\Ai$ and 
$\Aii$ after the projection of subsystem $\B$. 
Firstly, we take the transpose of $\rho_\A$ over the degrees of freedom 
in $\Aii$, 
\bea
\rho_\A^{T_2} 
&=& 
\frac{1}{\vert G_\A \vert} 
\sum_{g,g' \in g_\A}
\label{eq: rho_A proj T2}
\\ 
&\times& 
\left( \vphantom{\sum} 
  g_{\Ai} \vert 0_{\Ai} \rangle\langle 0_{\Ai} \vert g'_{\Ai} 
\right) 
\otimes 
\left( \vphantom{\sum} 
  g'_{\Aii} \vert 0_{\Aii} \rangle\langle 0_{\Aii} \vert g_{\Aii} 
\right) 
, 
\nonumber
\eea
and then we compute its second power, 
\bea
\left( \rho_\A^{T_2} \right)^2 
&=& 
\frac{1}{\vert G_\A \vert^2} 
\sum_{g,g' \in g_\A}
\sum_{\tg,\tg' \in g_\A}
\label{eq: rho_A proj T2 squared}
\\
&\times&
\left( \vphantom{\sum} 
  g_{\Ai} \vert 0_{\Ai} \rangle\langle 0_{\Ai} \vert g'_{\Ai} 
  \tg_{\Ai} \vert 0_{\Ai} \rangle\langle 0_{\Ai} \vert \tg'_{\Ai} 
\right) 
\nonumber \\
&\otimes&
\left( \vphantom{\sum}
  g'_{\Aii} \vert 0_{\Aii} \rangle\langle 0_{\Aii} \vert g_{\Aii} 
  \tg'_{\Aii} \vert 0_{\Aii} \rangle\langle 0_{\Aii} \vert \tg_{\Aii} 
\right)
. 
\nonumber 
\eea

Given $g$ and $g'$, we can change the summation variables 
$\tg \to \tg g'$ and $\tg' \to \tg' g$ (one-to-one mapping from $G_\A$ to 
$G_\A$, where we re-use the same labels before and after the change of 
variables for notational convenience): 
\bea
\left( \rho_\A^{T_2} \right)^2 
&=& 
\frac{1}{\vert G_\A \vert^2} 
\sum_{g,g' \in g_\A}
\sum_{\tg,\tg' \in g_\A}
\nonumber 
\\
&\times&
\left( \vphantom{\sum} 
  g_{\Ai} \vert 0_{\Ai} \rangle\langle 0_{\Ai} \vert \tg_{\Ai} 
  \vert 0_{\Ai} \rangle\langle 0_{\Ai} \vert \tg'_{\Ai} g_{\Ai}
\right) 
\nonumber \\
&\otimes&
\left( \vphantom{\sum}
  g'_{\Aii} \vert 0_{\Aii} \rangle\langle 0_{\Aii} \vert \tg'_{\Aii}
  \vert 0_{\Aii} \rangle\langle 0_{\Aii} \vert \tg_{\Aii} g'_{\Aii}
\right)
. 
\nonumber 
\eea
The two expectation values can thus be seen to impose the conditions 
$\tg_{\Ai} = \openone_{\Ai}$ and $\tg'_{\Aii} = \openone_{\Aii}$, or 
equivalently $\tg \in G_{\Aii}$ and $\tg' \in G_{\Ai}$: 
\bea
\left( \rho_\A^{T_2} \right)^2 
&=& 
\frac{1}{\vert G_\A \vert^2} 
\sum_{g,g' \in G_\A}
\sum_{\tg \in G_{\Aii}}
\sum_{\tg' \in G_{\Ai}}
\label{eq: rho_A proj T2 squared final}
\\
&& \!\!\!\!\!\!\!\!\!\!\!\!\!\!\!\!\!\!\!\!\!\!\!\!
\times
\left( \vphantom{\sum} 
  g_{\Ai} \vert 0_{\Ai} \rangle\langle 0_{\Ai} \vert \tg'_{\Ai} g_{\Ai}
\right) 
\otimes
\left( \vphantom{\sum}
  g'_{\Aii} \vert 0_{\Aii} \rangle\langle 0_{\Aii} \vert \tg_{\Aii} g'_{\Aii} 
\right)
. 
\nonumber 
\eea

Using Eq.~\eqref{eq: rho_A proj T2 squared final} we can proceed to compute 
the third power
\bea
\left( \rho_\A^{T_2} \right)^3 
&=& 
\frac{1}{\vert G_\A \vert^3} 
\sum_{g,g' \in G_\A}
\sum_{\tg \in G_{\Aii}}
\sum_{\tg' \in G_{\Ai}}
\sum_{h,h' \in G_\A}
\label{eq: rho_A proj T2 cubed}
\\
&& 
\times
\left( \vphantom{\sum} 
  g_{\Ai} \vert 0_{\Ai} \rangle\langle 0_{\Ai} \vert \tg'_{\Ai} g_{\Ai}
  h_{\Ai} \vert 0_{\Ai} \rangle\langle 0_{\Ai} \vert h'_{\Ai} 
\right) 
\nonumber \\
&& 
\otimes 
\left( \vphantom{\sum}
  g'_{\Aii} \vert 0_{\Aii} \rangle\langle 0_{\Aii} \vert \tg_{\Aii}  g'_{\Aii} 
  h'_{\Aii} \vert 0_{\Aii} \rangle\langle 0_{\Aii} \vert h_{\Aii} 
\right) 
. 
\nonumber 
\eea
Given $g' \in G_{\A}$ and $\tg \in G_{\Aii} \subset G_{\A}$, it is useful 
to redefine $h' \to \tg g' h'$. Similarly, we can redefine $h \to \tg' g h$. 
It is straightforward to show 
that both changes correspond to a trivial re-labelling the summation indices 
(one-to-one mappings of $G_\A$ onto $G_\A$): 
\bea
\left( \rho_\A^{T_2} \right)^3 
&=& 
\frac{1}{\vert G_\A \vert^3} 
\sum_{g,g' \in G_\A}
\sum_{\tg \in G_{\Aii}}
\sum_{\tg' \in G_{\Ai}}
\sum_{h,h' \in G_\A}
\nonumber
\\
&& \!\!\!\!\!\!\!\!\!\!\!\!
\times
\left( \vphantom{\sum} 
  g_{\Ai} \vert 0_{\Ai} \rangle\langle 0_{\Ai} \vert 
  h_{\Ai} \vert 0_{\Ai} \rangle\langle 0_{\Ai} \vert 
	\tg_{\Ai} g'_{\Ai} h'_{\Ai} 
\right) 
\nonumber \\
&& \!\!\!\!\!\!\!\!\!\!\!\!
\otimes 
\left( \vphantom{\sum}
  g'_{\Aii} \vert 0_{\Aii} \rangle\langle 0_{\Aii} \vert 
  h'_{\Aii} \vert 0_{\Aii} \rangle\langle 0_{\Aii} \vert 
	\tg'_{\Aii} g_{\Aii} h_{\Aii} 
\right) 
. 
\nonumber 
\eea
Notice that $\tg_{\Ai} = \openone_{\Ai}$ since $\tg \in G_{\Aii}$, and 
$\tg'_{\Aii} = \openone_{\Aii}$ since $\tg' \in G_{\Ai}$, by which the 
dependence on $\tg$ and $\tg'$ disappears (and they sum to an overall factor 
$\vert G_{\Ai} \vert \vert G_{\Aii} \vert$). 
Moreover, we can redefine $g' \to h' g'$ as well as $g \to h g$, 
thus transferring $h'$ ($h$) from the right hand side of the second 
(third) line in the equation above to the left hand side of the third 
(second) line. 
The two expectation values impose the conditions $h \in G_{\Aii}$ and 
$h' \in G_{\Ai}$, and therefore the dependence on $h$ and $h'$ also 
disappears (producing another overall factor 
$\vert G_{\Ai} \vert \vert G_{\Aii} \vert$). 

The result above thus simplifies to: 
\bea
\left( \rho_\A^{T_2} \right)^3 
&=& 
\frac{\vert G_{\Ai} \vert^2 \, \vert G_{\Aii} \vert^2}{\vert G_\A \vert^3} 
\sum_{g,g' \in G_\A}
\label{eq: rho_A proj T2 cubed final}
\nonumber
\\
&\times& 
\left( \vphantom{\sum} 
  g_{\Ai} \vert 0_{\Ai} \rangle\langle 0_{\Ai} \vert g'_{\Ai} 
\right) 
\otimes 
\left( \vphantom{\sum}
  g'_{\Aii} \vert 0_{\Aii} \rangle\langle 0_{\Aii} \vert g_{\Aii}
\right) 
\nonumber \\
&=& 
\frac{\vert G_{\Ai} \vert^2 \, \vert G_{\Aii} \vert^2}{\vert G_\A \vert^2} 
\;\rho_\A^{T_2}
. 
\eea
where the last equality was obtained by comparison with 
Eq.~\eqref{eq: rho_A proj T2}. 
We can finally combine our results to obtain that (for $n \geq 2$) 
\bea
\left( \rho_\A^{T_2} \right)^n 
&=& 
\left( 
  \frac{\vert G_{\Ai} \vert^2 \, \vert G_{\Aii} \vert^2}{\vert G_\A \vert^2} 
\right)^{k} 
\;\;
\left\{ 
  \begin{array}{ll}
	  \rho_\A^{T_2}  & \textrm{if $n = 2k+1$}
		\\ & \\
		\left( \rho_\A^{T_2} \right)^2 & \textrm{if $n = 2k+2$} . 
	\end{array}
\right. 
\nonumber
\eea
Given the fact that ${\rm Tr} \left( \rho_\A^{T_2} \right) = 1$ and 
${\rm Tr} \left[ \left( \rho_\A^{T_2} \right)^2 \right] = 1$, then 
\bea
\textrm{Tr}\left[ \left( \rho_\A^{T_2} \right)^n \right] 
&=& 
\left\{
\begin{array}{ll} 
  \left( 
    \frac{\vert G_{\Ai} \vert \, \vert G_{\Aii} \vert}{\vert G_\A \vert} 
  \right)^{n-1} 
  & 
	\textrm{if $n$ is odd} 
	\\ & \\
  \left( 
    \frac{\vert G_{\Ai} \vert \, \vert G_{\Aii} \vert}{\vert G_\A \vert} 
  \right)^{n-2} 
  &
	\textrm{if $n$ is even} 
\end{array}
\right.  
\label{eq: Tr rho_a^T2 n}
\eea
Following the replica approach proposed in Ref.~\onlinecite{Calabrese2012}, 
we see that the analytic continuation of Eq.~\eqref{eq: Tr rho_a^T2 n} 
for $n \to 1$ differs whether we follow the even or odd power sequence. 
The odd sequence tends to $1$, as expected for the trace of $\rho_\A^{T_2}$. 
The even sequence tends instead to the sum of the absolute values of the 
eigenvalues of $\rho_\A^{T_2}$ and it does not converge to $1$, thus 
signalling a non-vanishing entanglement negativity: 
\bea
\E &\equiv& \ln \| \rho_{\A}^{T_2} \|_1 
= 
\ln \frac{\vert G_\A \vert}{\vert G_{\Ai} \vert \, \vert G_{\Aii} \vert} 
= 
\ln 2 
. 
\eea
Once again, we obtain a direct measure of the topological entropy of the 
system \textit{without any boundary term or other contribution that scales 
with the size of the partition}. 
%
%

\subsection{\label{sec: generic state}
Choice of the projected state
           }
It is important to stress here that the results obtained thus far (both for 
the von Neumann entropy as well as the entanglement negativity) do not 
depend in any way on the specific choice of state $\vert 0_{\A} \rangle$. 
Therefore, they are also independent from the specific state that 
subsystem $\B$ is projected onto, provided it has a non-vanishing 
overlap with the ground state of the system. 

This holds true under the assumption that $\vert \phi_\B \rangle$ is a 
tensor product state of the $\sigma^z_i$ operators. 
Given the symmetry of the toric code upon exchanging $x \leftrightarrow z$ 
and stars with plaquettes, similar calculations and results 
hold true for projections on $\sigma^x_i$ tensor product states. 

What about more general projective measurements on $\B$? 
We can always express a generic state $\vert \phi_\B \rangle$ as a 
superposition of tensor product states in the $\sigma^z$ basis. 
The only tensor product states that are relevant to the present 
work are those which have non-vanishing overlap with the GS wavefunction 
of the toric code, i.e., of the form $\og_\B \vert 0_\B \rangle$ for some 
$\og \in G$. 
In App.~\ref{app: generic state} we discuss in detail the 
instructive example of a superposition of two such states. 
Here we report a summary of the results and relative discussion. 
We find that one ought to distinguish between states, say $\og$ and $\oog$, 
according to the whether {\it at least} one of the following conditions is 
satisfied: 
\bea
\exists h \in G_{\A} \; &:& \; h_{\Aii} \og_{\Aii} = \oog_{\Aii}
\label{eq: sup cond 1}\\
\exists h' \in G_{\A} \; &:& \; h'_{\Ai} \og_{\Ai} = \oog_{\Ai}
. 
\label{eq: sup cond 2}
\eea

If the answer is positive, one can then use the freedom in the choice of 
elements of subgroup $G_\A \subset G$ (which necessarily survives after 
projecting out $\B$) to interchange $\og$ with $\oog$, at least on 
one of the two components $\Ai$ or $\Aii$. 
Given that we are interested in measuring the entanglement between $\Ai$ and 
$\Aii$, the contributions from $\og$ and from $\oog$ are thence one and the 
same, and the results obtained for a single tensor product state remain 
unchanged. 

On the contrary, when neither condition is satisfiable, the two 
states $\og$ and $\oog$ give different contributions and we find 
that, in addition to the expected topological entropy, the von Neumann 
entropy and entanglement negativity pick up a contribution that depends on 
the choice of quantum superposition of $\og$ and $\oog$. 
In the case of a superposition 
$\alpha \og_\B \vert 0_\B \rangle + \beta \oog_\B \vert 0_\B \rangle$, 
we obtain in App.~\ref{app: generic state} the von Neumann entropy 
\bea
S_{\rm vN}^{(\Ai)} 
&=& 
\ln \frac{\vert G_\A \vert}{\vert G_{\Ai} \vert \vert G_{\Aii} \vert} 
- 
\vert \alpha \vert^2 \ln \vert \alpha \vert^2 
- 
\vert \beta \vert^2 \ln \vert \beta \vert^2 
, 
\nonumber\\
\label{eq: von Neumann superp}
\eea
and the logarithmic entanglement negativity 
\bea
\E 
&=& 
\ln \frac{\vert G_\A \vert}{\vert G_{\Ai} \vert \vert G_{\Aii} \vert} 
+ 
2 \ln 
\left( 
  \vert \alpha \vert
	+ 
	\vert \beta \vert
\right)
. 
\label{eq: negativity superp}
\eea
In addition to the usual topological contribution, a new term appears that 
directly depends on the weights of the states in the superposition. 
In Eq.~\eqref{eq: von Neumann superp} this takes the form of the classical 
entropy of mixing (recall that 
$\vert \alpha \vert^2 + \vert \beta \vert^2 = 1$); however, the presence of 
a similar term in the negativity confirms that it originates from quantum 
rather than classical correlations. 
It is interesting to notice that the additional contribution takes a similar 
but not identical form in $S_{\rm vN}^{(\Ai)}$, 
Eq.~\eqref{eq: von Neumann superp}, and in $\E$, 
Eq.~\eqref{eq: negativity superp}. 

A few comments are in order. Firstly, we note that the additional 
contribution is always positive. Therefore, the topological entropy $\gamma$ 
is a lower bound for the von Neumann entropy and entanglement negativity 
approaches proposed in this paper, with respect to the choice of projected 
state for subsystem $\B$. It may be possible to devise an appropriate 
minimisation routine on the latter to extract $\gamma$ without a priori 
knowledge of the suitable choice(s) for $\vert\phi_\B\rangle$. 

Secondly, when neither conditions in Eqs.~\eqref{eq: sup cond 1} 
and~\eqref{eq: sup cond 2}
are satisfied, we see from the calculation in App.~\ref{app: generic state}, 
e.g., Eq.~\eqref{eq: S_vN superp}, that the two states identified by 
$\og$ and $\oog$ contribute separately and additively to the entanglement 
measure (be it the von Neumann entropy or the negativity). 
Therefore, the projection of subsystem $\B$ onto the (equal amplitude) 
superposition of these states is akin to taking the trace over them. 
The corresponding additional contribution amounts to the logarithm of the 
number of such states. 
One can verify that Eqs.~\eqref{eq: sup cond 1} and~\eqref{eq: sup cond 2} 
are not satisfied only if $\og$ and $\oog$ differ by the action of a 
(product of) star operators acting at the boundaries $\partial_{\Ai}$ 
and $\partial_{\Aii}$. Their number scales exponentially in the length 
of the boundaries, and this recovers indeed the area law contribution to the 
entanglement measures upon tracing rather than projecting onto a given state 
of subsystem $\B$. 
%
%

\section{
Classical vs quantum 
entropy
        }
In general, the von Neumann entropy is sensitive to both classical and 
quantum correlations. On the contrary, the entanglement negativity is 
a measure of quantum correlations alone. 
For this reason, it is interesting to see how their behaviour differs, 
in the context of the approach discussed in the present paper, for 
a \textit{classical} topologically ordered system such as for instance the 
eight-vertex model~\cite{eight-vertex_refs} 
(see Ref.~\onlinecite{Castelnovo2007} for a calculation of the von Neumann 
entropy of this system). 

The classical eight-vertex model is a combinatorial problem of arrows on the 
bonds of the square lattice, with the hard constraint that the number of 
incoming arrows at every vertex is even. 
As discussed in App.~\ref{app: rho projected classical}, 
the density matrix of the toric code model, in the totally mixed limit where 
it is stripped of all its off-diagonal elements, reduces to the density 
matrix of the classical eight-vertex model. From this we obtain 
the (normalised) projected density matrix 
\bea
\rho^{\rm 8v}_\A 
&=& 
\frac{1}{\vert G_{\A} \vert} 
\sum_{g \in G_\A} 
g_\A \vert 0_\A \rangle \langle 0_\A \vert g_\A 
. 
\eea

Following the same steps used earlier in the von Neumann entropy calculation 
for the quantum case, we introduce $\rho^{\rm 8v}_{\Ai}$ by tracing over the 
degrees of freedom in $\Aii$, 
\bea
\rho^{\rm 8v}_{\Ai} 
&=& 
\frac{1}{\vert G_{\A} \vert} 
\sum_{g \in G_\A} 
g_{\Ai} \vert 0_{\Ai} \rangle \langle 0_{\Ai} \vert g_{\Ai} 
, 
\eea
and we compute 
\bea
\left( \rho^{\rm 8v}_{\Ai} \right)^2 
&=& 
\frac{1}{\vert G_{\A} \vert^2} 
\sum_{g,g' \in G_\A} 
\nonumber\\
&\times&
g_{\Ai} \vert 0_{\Ai} \rangle \langle 0_{\Ai} \vert g_{\Ai} 
g'_{\Ai} \vert 0_{\Ai} \rangle \langle 0_{\Ai} \vert g'_{\Ai} 
\nonumber\\ 
&=& 
\frac{\vert G_{\Aii} \vert}{\vert G_{\A} \vert^2} 
\sum_{g \in G_\A} 
g_{\Ai} \vert 0_{\Ai} \rangle \langle 0_{\Ai} \vert g_{\Ai} 
\nonumber \\ 
&=& 
\frac{\vert G_{\Aii} \vert}{\vert G_{\A} \vert} 
\;\rho^{\rm 8v}_{\Ai}
, 
\eea
where we made the change of variable $\tg = g' g$, and then realised that the 
expectation value imposes $\tg \in G_{\Aii}$ and the dependence on $\tg$ 
disappears. 

Iterating this identity, we obtain 
\bea
{\rm Tr} \left[ \left( \rho^{\rm 8v}_{\Ai} \right)^2 \right] 
&=& 
\left( \frac{\vert G_{\Aii} \vert}{\vert G_{\A} \vert} \right)^{n-1} 
, 
\eea
and 
\bea
S^{(\Ai)}_{\rm vN} 
&=& 
- \ln \left( \frac{\vert G_{\Aii} \vert}{\vert G_{\A} \vert} \right)
= 
\ln \left( 2^{N^{(s)}_{\Ai}} \right) 
+ 
\ln \left( 2^{m_{\B} - 1} \right) 
, 
\nonumber 
\eea
where we used again the fact that 
$\vert G_{\Aii} \vert = 2^{N^{(2)}_{\Aii}}$ 
and $\vert G_{\A} \vert = 2^{N^{(2)}_{\A} + m_{\B} - 1}$ 
($m_{\B} = 2$ for the choice of partitions in Fig.~\ref{fig: partitions}). 
The first contribution is the expected extensive term (recall that the 
von Neumann entropy becomes a measure of the statistical mechanical entropy 
for classical systems) whereas the second (order one) contribution 
is a signature of the classical topologically ordered nature of the system. 

Whereas the proposed approach via projecting the degrees of freedom 
in $\B$ no longer leads to a distillation of the sole topological 
contribution, we observe nonetheless a signature of the classical 
topological entropy of the system. 

Let us contrast this result with the negativity calculation for the same 
system. Instead of tracing over $\Aii$, we take the transpose, which 
however leaves a purely diagonal density matrix unchanged, 
$\left( \rho^{\rm 8v}_\A \right)^{T_2} \equiv \rho^{\rm 8v}_\A$. 
The calculation of its square becomes therefore straightforward, 
\bea
\left( \rho^{\rm 8v}_\A \right)^2 
&=& 
\frac{1}{\vert G_{\A} \vert^2} 
\sum_{g,g' \in G_\A} 
g_\A \vert 0_\A \rangle \langle 0_\A \vert g_\A 
g'_\A \vert 0_\A \rangle \langle 0_\A \vert g'_\A 
\nonumber \\
&=& 
\frac{1}{\vert G_{\A} \vert} 
\rho^{\rm 8v}_\A
, 
\eea
since the expectation value selects $g'$ uniquely by imposing the condition 
$g' = g$. From this result we see that 
$
{\rm Tr} \left[ \left( \rho^{\rm 8v}_\A \right)^n \right] 
= 
1/\vert G_{\A} \vert^{n-1} 
$, 
which has the same analytic continuation for $n$ even or odd and therefore 
the negativity vanishes, as expected for a system with purely classical 
correlations. 

In summary, the calculation of the von Neumann entropy 
for a classical system contains a contribution due to the classical 
topological entropy, which is lost (as expected) in the negativity 
calculation. 
The clear difference in behaviour of the proposed 
von Neumann and negativity measures between quantum and classical systems 
can thus be used to distinguish between the two instances of topological 
order. 
%
%

\section{
Conclusions
        }
Obtaining the topological entropy of a system from analytical 
calculations or numerical simulations is in general a tall order, since 
it appears in measures of entanglement as an order one correction to 
leading terms that scale with the size of the system. 

In this paper we have shown that one can distil the topological entropy alone 
using von Neumann entropy or entanglement negativity measures 
where the conventional trace over part of the system is replaced with a 
projection. 
The combination of such projection (the state of which needs not be known) 
with an appropriate choice of partitions allows to remove the unwanted 
boundary terms. 
The topological entropy of a system can thus be obtained via a 
single measurement, without the conventional need for 
elaborate subtraction schemes or extrapolation as a function of system size. 


Given that the entanglement entropy computed from $\rho_{\Ai}$ in 
Eq.~\eqref{eq: rho_A1} is devoid of a boundary term, and yet not topologically 
trivial, it will be interesting to compare the corresponding entanglement 
spectrum with the one from the reduced density matrix obtained by tracing 
over both $\Aii$ and $\B$. 
This is however beyond the scope of the present paper. 

In the paper, we also discussed the conditions that the 
projected wavefunction for $\B$ has to satisty in order for our results to 
hold. For instance, any tensor product state in the $\sigma^z$ or $\sigma^x$ 
basis is suitable. However, superpositions thereof require appropriate 
relations between the states involved, which are summarised by 
Eq.~\eqref{eq: sup cond 1} and Eq.~\eqref{eq: sup cond 2}. 
When the conditions are violated, the von Neumann entropy and entanglement 
negativity acquire additional contributions on top of the expected 
topological entropy, which depend on the amplitudes of the states in the 
superposition. 
The additional contributions are always positive; 
it may therefore be possible to combine an appropriate minimisation 
procedure over the projected state with 
the approach in this paper to find suitable states where these contributions 
vanish. 

We further showed that similar results apply to instances of classical 
topological order, provided that we use the von Neumann entanglement entropy. 
In this case, the topological correlations are purely classical in nature 
and the negativity vanishes identically. 
The combined use of both the von Neumann entropy and of the entanglement 
negativity allows thus to distinguish and characterise classical and quantum 
topolgical correlations. 

Here we considered only the toric code at zero-temperature in two dimensions. 
From the results in Refs.~\onlinecite{Castelnovo2007finT,Castelnovo2008finT}, 
one expects the topological entropy to vanish in this system at any finite 
temperature in the thermodynamic limit. 
On the contrary, the toric code in 3D reduces to a classical 
$\mathbb{Z}_2$ gauge theory at finite temperature, which remains 
topologically ordered albeit only classically up to a finite temperature 
phase transition. 
Using the approach discussed in this paper and the results from both 
the von Neumann entropy and the entanglement negativity, one ought to 
observe that the former survives up to the 
transition (as a measure of classical 
topological entropy), 
whereas the latter vanishes at any finite temperature (as a measure of 
quantum topological entropy only). 


This is to be contrasted with, say, the 3D toric code at zero temperature 
in presence of a magnetic field, where both measures of the topological 
entropy survive as the field is increased, up to a quantum phase 
transition at finite field. 

It will be interesting to see how the finite size behaviour of the 
negativity at finite field / finite temperature differs, and in particular 
whether the finite size behaviour in the negativity calculations of the 
quantum topological entropy bears a signature of the zero temperature 
topological order at sufficiently small but finite temperatures 
(cf. for example the finite size behaviour of the entanglement 
entropy in Refs.~\onlinecite{Castelnovo2007finT,Castelnovo2008finT}). 
The ability to compute only the topological entropy without boundary terms 
scaling with the size of the system may help to study these different 
scenarios. 

Similar calculations could be extended to other 3D systems that 
have been recently argued to exhibit $\mathbb{Z}_2$ quantum spin liquid 
behaviour at finite temperature~\cite{Lee2013b,Kimchi2013,Nasu2013}. 
This would allow to test whether quantum topological order in such 
systems does indeed survive at finite temperature (as measured by the 
topological contribution to the entanglement), or it vanishes 
in the thermodynamic limit as is the case for the 3D toric code. 

To some extent the toric code is a rather special example of topological 
order with precisely `zero-ranged' local correlations. 
It will be interesting to study extensions of these calculations of the 
topological entropy, perhaps numerically, to more complex examples of 
topologically non-trivial states. 

One could investigate perturbations of the toric code introduced via 
stochastic matrix form decomposition~\cite{Castelnovo2008}, where the GS 
wavefunction is known exactly throughout the phase diagram. 
These perturbations introduce finite correlations and eventually drive the 
system across a so-called conformal critical point. 
It may also be possible to study the behaviour of the negativity at such 
critical points by means of conformal field theoretic 
techniques~\cite{Fradkin2006,Castelnovo2007,Oshikawa2010}. 
This work could lead the way to the even more interesting 
challenge of applying this approach to quantum Hall states and other 
topologically ordered phases of matter. 
%

As a closing remark, one should remember that a non-vanishing topological 
entropy per se is not evidence of topological order but rather an indication 
that the system \textit{can} exhibit topological order. 
For instance, if a non-local term was added to the Hamiltonian which 
selected uniquely one of the degenerate ground states, the results presented 
in this paper would remain unchanged. 
In this respect, we notice that there is a difference between the choice of 
partitions in the top and bottom panels in Fig.~\ref{fig: partitions}. The 
bottom panel represents a \textit{local} test of the ability of the system 
to support topological order, since it only looks at topological 
correlations within the outer boundary of $\Aii$. 
%
%

\section*{
Acknowledgments
         }
We are grateful to G.~Vidal and C.~Chamon for useful discussions. 
This work was supported by EPSRC Grant EP/K028960/1, and in part by the 
Helmholtz Virtual Institute ``New States of Matter and Their Excitations'' 
and by the EPSRC NetworkPlus on ``Emergence and Physics far from Equilibrium''. 
%
%

\appendix
%
%

\section{\label{app: rho projected}
Projected $\rho_\A$ (quantum)
        }
Let us consider the density matrix of the system prepared in the GS in 
Eq.~\eqref{eq: TC GS}, 
\beq
\rho = \vert \psi_0 \rangle \langle \psi_0 \vert 
= \frac{1}{\vert G \vert} 
\sum_{g,g' \in G} 
g \vert 0 \rangle \langle 0 \vert g' 
, 
\eeq
and compute the projected density matrix 
\beq
\rho_\A = \langle \phi_\B \vert \rho \vert \phi_\B \rangle 
\eeq
where 
$\vert \phi_\B \rangle = \og_\B \vert 0_\B \rangle$, for some $\og \in G$. 

It is convenient to redefine $g \to g \og$ and 
$g' \to g' \og$, which are 
one-to-one mappings in $G$. Using the fact that 
$(\og)^2 = \openone$ (and equivalently 
$(\og_\B)^2 = \openone_\B$), 
the projected density matrix can then be written as 
\bea
\rho_\A 
&=& 
\frac{1}{\vert G \vert} 
\sum_{g,g' \in G} 
g_\A \og_\A \vert 0_\A \rangle \langle 0_\A \vert \og_\A g'_\A 
\nonumber \\
&\times& 
\langle 0_\B \vert g_\B \vert 0_\B \rangle 
\langle 0_\B \vert g'_\B \vert 0_\B \rangle 
, 
\eea
where  we introduced the notation 
$\vert 0 \rangle = \vert 0_{\A} \rangle \otimes \vert 0_{\B} \rangle$ 
and $g = g_{\A} \otimes g_{\B}$. 
The two expectation values impose that $g_\B = g'_\B = \openone_\B$, 
i.e., $g,g' \in G_\A \subset G$. 
At the same time, we can redefine the reference state 
$\vert 0 \rangle \to \og \vert 0 \rangle$ (recall that the choice of 
reference state in Eq.~\eqref{eq: TC GS} was arbitrary) and we arrive at 
the expression 
\bea
\rho_\A 
&=& 
\frac{1}{\vert G \vert} 
\sum_{g,g' \in G_\A} 
g_\A \vert 0_\A \rangle \langle 0_\A \vert g'_\A 
. 
\label{eq: rho_A proj Appendix}
\eea
For convenience, we further normalise $\rho_\A$ by replacing the factor 
$1/\vert G \vert$ with $1/\vert G_\A \vert$. 


\section{\label{app: S_vN projected}
Entanglement entropy of the projected $\rho_\A$
        }
It is straightforward to show that the von Neumann entropy of subsystem $\A$ 
vanishes once $\B$ has been projected to a given state 
$\og_\B \vert 0_\B \rangle$. 
In order to obtain the von Neumann entropy of $\rho_\A$ in 
Eq.~\eqref{eq: rho_A proj Appendix} (equivalently, Eq.~\eqref{eq: rho_A proj} 
in the main text), we need to compute 
\bea
\rho_\A^2 
&=& 
\frac{1}{\vert G_\A \vert^2} 
\sum_{g,g' \in G_\A} 
\sum_{\tg,\tg' \in G_\A} 
g_\A \vert 0_\A \rangle \langle 0_\A \vert g'_\A 
\tg_\A \vert 0_\A \rangle \langle 0_\A \vert \tg'_\A 
. 
\nonumber
\eea
The product $g' \tg$ is a generic element of $G_\A$, since both $g'$ and 
$\tg$ belong to the same group. Therefore, the corresponding expectation 
value fixes uniquely the product to the identity $\openone$, and we are 
left with a free summation over the elements of $G_\A$ (i.e., a factor of 
$\vert G_\A \vert$): 
\bea
\rho_\A^2 
&=& 
\frac{1}{\vert G_\A \vert} 
\sum_{g \in G_\A} 
\sum_{\tg' \in G_\A} 
g_\A \vert 0_\A \rangle \langle 0_\A \vert \tg'_\A 
= \rho_\A 
. 
\nonumber
\eea
Iterating this identity, one finds that $\rho_\A^n = \rho_\A$, 
${\rm Tr} (\rho_\A^n) = 1$, and 
$S_{\rm vN}^{(\A)} = - \lim_{n \to 1} \partial_n [{\rm Tr} (\rho_\A^n)] = 0$. 


\section{\label{app: rho projected classical}
Projected $\rho_\A$ (classical)
        }
The classical eight-vertex model is a combinatorial problem of arrows on the 
bonds of the square lattice, with the hard constraint that the number of 
incoming arrows at every vertex is even (counting $0$ as an even number). 
Taking advantage of the bipartite nature of the lattice, we can define 
arrows going from sublattice A to sublattice B as positive spins, and all 
others are negative. This establishes a 1-to-1 mapping between eight-vertex 
configurations and $\sigma^z$ tensor product states that minimise the energy 
of the plaquette term in the toric code Hamiltonian, 
Eq.~\eqref{eq: TC Hamiltonian}. 
All eight-vertex configurations can be obtained from a reference configuration, 
say the spin polarized $\vert 0 \rangle$, by acting with elements of $G$. 
The eight-vertex model represents an instance of a classical topologically 
ordered system~\cite{Castelnovo2007}. 

The density matrix of the toric code model, in the totally mixed limit where 
it is stripped of all its off-diagonal elements, 
\beq
\rho = \frac{1}{\vert G \vert} 
\sum_{g \in G} 
g \vert 0 \rangle \langle 0 \vert g 
, 
\eeq
reduces thus to the density matrix of the classical eight-vertex model. 

Let us compute the projected density matrix in this case, 
$\rho_\A = \langle \phi_\B \vert \rho \vert \phi_\B \rangle$, 
with 
$\vert \phi_\B \rangle = \og_\B \vert 0_\B \rangle$, for some 
$\og \in G$. 
Using once again the change of summation variable $g \to g \og$, 
the projected density matrix can then be written as 
\bea
\rho_\A 
&=& 
\frac{1}{\vert G \vert} 
\sum_{g \in G} 
g_\A \og_\A \vert 0_\A \rangle \langle 0_\A \vert \og_\A g_\A 
\nonumber \\
&\times& 
\langle 0_\B \vert g_\B \vert 0_\B \rangle 
\langle 0_\B \vert g_\B \vert 0_\B \rangle 
, 
\eea
where the two expectation values impose that $g_\B = \openone_\B$, 
i.e., $g \in G_\A$. Once we redefine the reference state 
$\vert 0 \rangle \to \og \vert 0 \rangle$ (recall that the choice of 
reference state in Eq.~\eqref{eq: TC GS} was indeed arbitrary), we arrive at 
the expression 
\bea
\rho_\A 
&=& 
\frac{1}{\vert G \vert} 
\sum_{g \in G_\A} 
g_\A \vert 0_\A \rangle \langle 0_\A \vert g_\A 
. 
\eea
As before, we finally normalise $\rho_\A$ by replacing the factor 
$1/\vert G \vert$ with $1/\vert G_\A \vert$. 


\section{\label{app: alternative negativity}
Negativity of the partitions in Fig.~\ref{fig: partitions}
        }
The entanglement negativity for the partitions shown in 
Fig.~\ref{fig: partitions}, after tracing over $\B$ and transposing $\Aii$, 
was computed in Refs.~\onlinecite{Lee2013,Castelnovo2013}. 
Here we present a more streamlined version of the calculation in 
Ref.~\onlinecite{Castelnovo2013}, which is at the basis of the 
results presented in the main text of the paper after projecting the degrees 
of freedom in $\B$. 

We start with the density matrix of the system prepared in the GS in 
Eq.~\eqref{eq: TC GS}, 
\beq
\rho 
= 
\vert \psi_0 \rangle \langle \psi_0 \vert 
= 
\frac{1}{\vert G \vert} 
\sum_{g,g' \in G} 
g \vert 0 \rangle \langle 0 \vert g' 
, 
\eeq
and take the trace over the degrees of freedom in $\B$, 
\beq
\rho_\A 
= 
\frac{1}{\vert G \vert} 
\sum_{g,g' \in G} 
g_\A \vert 0_\A \rangle \langle 0_\A \vert g'_\A 
\; \langle 0_\B \vert g'_\B g_\B \vert 0_\B \rangle 
. 
\eeq
It is then convenient to redefine $g' \to \tg = g' g$ (one-to-one mapping 
in $G$) and replace the summation over $g'$ with a summation over $\tg$. 
The expectation value then restricts $\tg$ to act as the identity on 
subsystem $\B$, i.e., $\tg \in G_{\A}$: 
\beq
\rho_\A 
= 
\frac{1}{\vert G \vert} 
\sum_{g \in G, \: \tg \in G_{\A}} 
g_\A \vert 0_\A \rangle \langle 0_\A \vert g_\A \tg_\A 
. 
\eeq
Notice that this expression differs from the projected $\rho_\A$ 
Eq.~\eqref{eq: rho_A proj} in that $g$ spans the whole group 
$G$ rather than its subgroup $G_\A$. This difference will however play a 
crucial role in the calculation of the negativity. 

After taking the transpose over the degrees of freedom in $\Aii$, 
\bea
\rho_\A^{T_2} 
&=& 
\frac{1}{\vert G \vert} 
\sum_{g \in G, \: \tg \in G_{\A}} 
\\
&\times& 
\left( 
  g_{\Ai} \vert 0_{\Ai} \rangle \langle 0_{\Ai} \vert g_{\Ai} \tg_{\Ai} 
\right) 
\nonumber \\
&\otimes&
\left( 
  g_{\Aii} \tg_{\Aii} \vert 0_{\Aii} \rangle \langle 0_{\Aii} \vert g_{\Aii} 
\right) 
. 
\nonumber
\eea
we compute the second power of $\rho_\A^{T_2}$: 
\bea
\left( \rho_\A^{T_2} \right)^2
&=& 
\frac{1}{\vert G \vert^2} 
\sum_{g_{1} \in G} 
\sum_{\tg_{1} \in G_{\A}} 
\sum_{g \in G} 
\sum_{\tg \in G_{\A}} 
\\
&& \!\!\!\!\!\!\!\!\!\!\!\!\!\!\!\!\!
\times 
\left( 
  g_{1\Ai} \vert 0_{\Ai} \rangle 
	\langle 0_{\Ai} \vert 
	  g_{1\Ai} \tg_{1\Ai} g_{\Ai} 
	\vert 0_{\Ai} \rangle 
	\langle 0_{\Ai} \vert g_{\Ai} \tg_{\Ai} 
\right) 
\nonumber \\
&& \!\!\!\!\!\!\!\!\!\!\!\!\!\!\!\!\!
\otimes
\left( 
  g_{1\Aii} \tg_{1\Aii} \vert 0_{\Aii} \rangle 
	\langle 0_{\Aii} \vert 
	  g_{1\Aii} g_{\Aii} \tg_{\Aii} 
	\vert 0_{\Aii} \rangle 
	\langle 0_{\Aii} \vert g_{\Aii} 
\right) 
. 
\nonumber
\eea

One can replace the summation over $g$ with a summation over 
$g_{2} = g_{1} g \in G$, upon substituting $g = g_{1} g_{2}$ in 
the expression above (one-to-one mapping in $G$, given $g_{1}$). 
If we also relabel $\tg \equiv \tg_{2}$, we obtain the following expression: 
\bea
\left( \rho_\A^{T_2} \right)^2
&=& 
\frac{1}{\vert G \vert^2} 
\sum_{g_{1},g_{2} \in G} 
\sum_{\tg_{1},\tg_{2} \in G_{\A}} 
\\
&& \!\!\!\!\!\!\!\!\!\!\!\!\!\!\!\!\!
\times
\left( 
  g_{1\Ai} \vert 0_{\Ai} \rangle 
	\langle 0_{\Ai} \vert 
	  g_{2\Ai} \tg_{1\Ai} 
	\vert 0_{\Ai} \rangle 
	\langle 0_{\Ai} \vert g_{1\Ai} g_{2\Ai} \tg_{2\Ai} 
\right) 
\nonumber \\
&& \!\!\!\!\!\!\!\!\!\!\!\!\!\!\!\!\!
\otimes
\left( 
  g_{1\Aii} \tg_{1\Aii} \vert 0_{\Aii} \rangle 
	\langle 0_{\Aii} \vert 
	  g_{2\Aii} \tg_{2\Aii} 
	\vert 0_{\Aii} \rangle 
	\langle 0_{\Aii} \vert g_{1\Aii} g_{2\Aii}
\right) 
. 
\nonumber
\eea

Similar considerations lead to the third power of $\rho_\A^{T_2}$, from which 
one can recognise the general pattern: 
\begin{widetext}
\bea
\left( \rho_\A^{T_2} \right)^n
&=& 
\frac{1}{\vert G \vert^n} 
\sum_{g_{1},\ldots,g_{n} \in G} \; 
\sum_{\tg_{1},\ldots,\tg_{n} \in G_{\A}} 
\\
&\times& 
\left[ 
  g_{1\Ai} \vert 0_{\Ai} \rangle 
  \prod_{\ell = 2}^n
	\langle 0_{\Ai} \vert 
	  g_{\ell\Ai} \tg_{(\ell-1)\Ai} 
	\vert 0_{\Ai} \rangle 
	\langle 0_{\Ai} \vert 
	\left( \prod^n_{\ell = 1} g_{\ell\Ai} \right) \tg_{n\Ai} 
\right] 
\nonumber \\
&\otimes&
\left[ 
  g_{1\Aii} \tg_{1\Aii} \vert 0_{\Aii} \rangle 
	\prod_{\ell = 2}^{n} 
	\langle 0_{\Aii} \vert 
	  g_{\ell\Aii} \tg_{\ell\Aii} 
	\vert 0_{\Aii} \rangle 
	\langle 0_{\Aii} \vert \prod^n_{\ell = 1} g_{\ell\Aii} 
\right] 
. 
\nonumber
\eea
%
%

Upon taking the trace, one notices that the dependence on $g_{1}$ 
disappears, thus resulting in an overall factor $\vert G \vert$: 
%
%
\bea
{\rm Tr} \left[ \left( \rho_\A^{T_2} \right)^n \right] 
&=& 
\frac{1}{\vert G \vert^{n-1}} 
\sum_{g_{2},\ldots,g_{n} \in G} \; 
\sum_{\tg_{1},\ldots,\tg_{n} \in G_{\A}} 
\label{eq: Tr rho_A^T2 pow n middle}
\\
&\times& 
  \prod_{\ell = 2}^n
	\langle 0_{\Ai} \vert 
	  g_{\ell\Ai} \tg_{(\ell-1)\Ai} 
	\vert 0_{\Ai} \rangle 
	\langle 0_{\Ai} \vert 
	\left( \prod^n_{\ell = 2} g_{\ell\Ai} \right) \tg_{n\Ai} 
  \vert 0_{\Ai} \rangle 
\nonumber \\
&\times&
	\prod_{\ell = 2}^{n} 
	\langle 0_{\Aii} \vert 
	  g_{\ell\Aii} \tg_{\ell\Aii} 
	\vert 0_{\Aii} \rangle 
	\langle 0_{\Aii} \vert 
	  \left( \prod^n_{\ell = 2} g_{\ell\Aii} \right) 
    \tg_{1\Aii} 
	\vert 0_{\Aii} \rangle 
. 
\nonumber
\eea
\end{widetext}
Moreover, the first $n-1$ expectation values in the second and third rows 
of Eq.~\eqref{eq: Tr rho_A^T2 pow n middle} impose that 
$g_{\ell\Ai} = \tg_{(\ell-1)\Ai}$ and 
$g_{\ell\Aii} = \tg_{\ell\Aii}$, for $\ell=2,\ldots,n$. 
Using these identities in the last expectation value in each row, the 
expression above reduces to 
%
%
\bea
&&
{\rm Tr} \left[ \left( \rho_\A^{T_2} \right)^n \right] 
= 
\frac{1}{\vert G \vert^{n-1}} 
\sum_{g_{2},\ldots,g_{n} \in G} \; 
\sum_{\tg_{1},\ldots,\tg_{n} \in G_{\A}} 
\\
&&\times 
  \prod_{\ell = 2}^n
	\langle 0_{\Ai} \vert 
	  g_{\ell\Ai} \tg_{(\ell-1)\Ai} 
	\vert 0_{\Ai} \rangle 
	\langle 0_{\Ai} \vert 
	  \prod^n_{\ell = 1} \tg_{\ell\Ai} 
  \vert 0_{\Ai} \rangle 
\nonumber \\
&&\times
	\prod_{\ell = 2}^{n} 
	\langle 0_{\Aii} \vert 
	  g_{\ell\Aii} \tg_{\ell\Aii} 
	\vert 0_{\Aii} \rangle 
	\langle 0_{\Aii} \vert 
	  \prod^n_{\ell = 1} \tg_{\ell\Aii} 
	\vert 0_{\Aii} \rangle 
. 
\nonumber
\eea
%
%
The last expectation values on each row, combined, impose that the 
product $\prod^n_{\ell = 1} \tg_{\ell}$ acts as the identity on $\A$, 
which fixes uniquely one of the $\tg_{\ell}$, since they are all elements 
of $G_\A$. 
The remaining $2(n-1)$ expectation values impose 
\beq
g_{\ell\A} = \tg_{(\ell-1)\Ai} \otimes \tg_{\ell\Aii} 
\label{eq: n-1 conditions}
\eeq
for all $\ell = 2,\ldots,n$. 
If this condition can be met, then the only freedom left in the choice of 
$g_{\ell} \in G$ is the multiplication by a generic element in $G_\B$, 
which results in a factor $\vert G_\B \vert$ upon summation over $g_{\ell}$. 

Whether the chosen density matrix and partitions have a vanishing or 
non-vanishing negativity is entirely 
dependent on whether there exists at least one element in $G$ that can 
satisfy the condition in Eq.~\eqref{eq: n-1 conditions}, for all 
$\ell = 2,\ldots,n$. 
If the answer is positive, then the trace reduces to 
\bea
{\rm Tr} \left[ \left( \rho_\A^{T_2} \right)^n \right] 
&=& 
\frac{\vert G_\A \vert^{n-1} \vert G_\B \vert^{n-1}}{\vert G \vert^{n-1}} 
\nonumber
\eea
and the negativity vanishes identically (no difference between the even and 
odd $n$ analytic continuations). 

Let us consider the condition in Eq.~\eqref{eq: n-1 conditions} in the 
context of the partitions in Fig~\ref{fig: partitions}. 
The group $G_\A$ can be decomposed as the product of three groups, 
$G_{\Ai}$, $G_{\Aii}$, and $G_{\Ai\Aii}$, where $G_{\Ai}$ and $G_{\Aii}$ 
act only on $\Ai$ and $\Aii$, respectively, 
and $G_{\Ai\Aii}$ is defined as the quotient group 
$G_\A / (G_{\Ai} G_{\Aii})$. 
Each element $\tg_{\ell} \in G_\A$ can correspondingly be uniquely 
decomposed as the product 
$\tg^{(1)}_{\ell} \, \tg^{(2)}_{\ell} \, \tg^{(12)}_{\ell}$ 
of three elements from each of the subgroups. 

In this notation, the condition in Eq.~\eqref{eq: n-1 conditions} can be 
written as 
\bea
g_{\ell\A} 
&=& 
\tg^{(1)}_{(\ell-1)\Ai} \tg^{(12)}_{(\ell-1)\Ai} 
\otimes 
\tg^{(2)}_{\ell\Aii} \tg^{(12)}_{\ell\Aii} 
\nonumber \\
&=& 
\left( 
  \tg^{(1)}_{(\ell-1)\Ai} 
  \otimes 
  \tg^{(2)}_{\ell\Aii} 
\right)
\left(
  \tg^{(12)}_{(\ell-1)\Ai} 
  \otimes 
  \tg^{(12)}_{\ell\Aii} 
\right)
. 
\eea
The first of the two factors is trivial since it is nothing but the 
product of $\tg^{(1)}_{\ell-1} \in G_{\Ai}$ times 
$\tg^{(2)}_{\ell} \in G_{\Aii}$, 
which is an element of $G$ and thus can be matched by an appropriate 
choice of $g_{\ell}$. 
On the other hand, the second product 
$\tg^{(12)}_{(\ell-1)\Ai} \otimes \tg^{(12)}_{\ell\Aii}$ 
is not obviously the action on $\A$ of an element of $G$, and one ought to 
consider it with care. 

With the choice in Fig.~\ref{fig: partitions}, the group $G_{\Ai\Aii}$ 
has only two elements: the identity, and the product $\theta$ of all star 
operators that act on at least one spin in $\Bi$. One can see that this 
element $\theta \in G$  acts simultaneously on $\Ai$ and $\Aii$ and cannot 
be written as a product of star operators acting solely (and separately) 
on spins in $\Ai$ and $\Aii$. 
With two choices each for $\tg^{(12)}_{\ell-1}$ and $\tg^{(12)}_{\ell}$, the 
non trivial option that requires consideration is when one of them is equal 
to $\theta$ and the other is the identity. In this case, we need to ask 
whether an element $g_{\ell} \in G$ exists such that 
$g_{\ell\Ai} = \theta_{\Ai}$ and $g_{\ell\Aii} = \openone_{\Aii}$ 
(or vice versa). 

The action of $\theta_{\Ai}$ is to flip all the spins in $\Ai$ that belong 
to stars at the boundary with $\Bi$. The product of all the corresponding 
boundary star operators is clearly an element $g^* \in G$ that acts on 
the system by flipping all the spins flipped by $\theta_{\Ai}$ as well as 
some spins in $\Bi$. 
Therefore, $g^{*}_{\A} = \theta_{\Ai} \otimes \openone_{\Aii}$ and 
the condition in Eq.~\eqref{eq: n-1 conditions} is satisfied. 
Since this constructive argument applies in general for the choice of 
partitions in Fig.~\ref{fig: partitions}, we conclude that 
the negativity vanishes identically. 


\section{\label{app: generic state}
Projection onto a generic state
        }
In the main text we considered the case where subsystem $\B$ 
is projected onto one of the $\sigma^z$ tensor product states. 
Here we discuss the case of a generic state, which can always be written as 
a superposition of $\sigma^z$ tensor product states. 
We show that the 
results obtained earlier hold only for a special class of superpositions, 
and otherwise the entanglement entropy becomes a function of both the 
topological contribution \textit{and} of the amplitudes in the 
projected wavefunction. 

Notice that, for a generic superposition 
\bea
\vert \phi_\B \rangle 
&=& 
\sum_{\{ \sigma^z_i \}_\B} 
  \alpha\left( \{ \sigma^z_i \}_\B \right)
	\vert \{ \sigma^z_i \}_\B \rangle 
, 
\eea
the only states that give a non-vanishing contribution to the 
projection of the density matrix of the system prepared in the state 
$\vert\psi_0\rangle$, Eq.~\eqref{eq: TC GS}, are those where 
$\vert \{ \sigma^z_i \}_\B \rangle = \og_\B \vert 0_\B \rangle$ 
for some $\og \in G$. 
For simplicity, we focus on the case of a coherent superposition between 
only two such states, which is sufficient for the purpose of this discussion: 
\bea
\vert \phi_\B \rangle 
&=& 
\alpha \og_{\B} \vert 0_\B \rangle + \beta \oog_{\B} \vert 0_\B \rangle 
. 
\eea
The state is assumed normalised, 
$\vert \alpha \vert^2 + \vert \beta \vert^2 = 1$. 

The (normalised) projected density matrix can then be written as 
\bea
\rho_{\A} 
&=& 
\frac{1}{\N}
\sum_{g, g' \in G} 
  \left( \vphantom{\sum}
    \alpha^* \langle 0_\B \vert \og_\B
    +
    \beta^* \langle 0_\B \vert \oog_\B 
	\right)
	g_\B \vert 0_\B \rangle 
\label{eq: rho proj 2 states}
\\
&\times& 
	\langle 0_\B \vert g'_\B 
  \left( \vphantom{\sum}
  	\alpha \og_\B \vert 0_\B \rangle
	  +
		\beta \oog_\B \vert 0_\B \rangle
	\right)
  \: 
  g_\A \vert 0_\A \rangle \langle 0_\A \vert g'_\A 
. 
\nonumber
\eea
Using a simplified notation to save space (the state $0_\B$ and the 
subscripts $\B$ are understood), the two matrix elements can be expanded as 
\bea
&&
\left( \vphantom{\sum}
  \alpha^* \langle \og \vert
  +
  \beta^* \langle \oog \vert
\right) 
\vert g \rangle 
\langle g' \vert
\left( \vphantom{\sum}
	\alpha \vert \og \rangle
  +
	\beta \vert \oog \rangle
\right)
\nonumber\\
&& \qquad
=
\left[
  \alpha^* \langle \og g \rangle
  +
  \beta^* \langle \oog g \rangle
\right]
\left[
  \alpha \langle g' \og \rangle
  +
  \beta \langle g' \oog \rangle
\right]
\nonumber\\
&& \qquad
=
\vert \alpha \vert^2 \langle \og g \rangle \langle g' \og \rangle
+ 
\alpha^* \beta \langle \og g \rangle \langle g' \oog \rangle
\nonumber \\ 
&& \qquad 
+ 
\beta^* \alpha \langle \oog g \rangle \langle g' \og \rangle
+ 
\vert \beta \vert^2 \langle \oog g \rangle \langle g' \oog \rangle
. 
\eea
We can then perform the summation in Eq.~\eqref{eq: rho proj 2 states} 
separately for each of the four terms, after changing variables so as to 
reduce the expectation values to the form 
$\langle g \rangle \langle g' \rangle$ (for instance for the first term 
proportional to $\vert \alpha \vert^2$ one needs to redefine 
$g \to \og g$ and $g' \to \og g'$). 
The expectation values impose $g,g' \in G_{\A}$, and we obtain 
the expression 
\bea
\rho_{\A} 
&=& 
\frac{1}{\N} 
\sum_{g, g' \in G_\A} 
  \left[ \vphantom{\sum} 
    \vert \alpha \vert^2 \:
		g_\A \og_\A \vert 0_\A \rangle \langle 0_\A \vert \og_\A g'_\A 
  \right.
\label{eq: rho proj 2 states bis}
\\
    && \qquad\qquad +
    \alpha^* \beta \:
		g_\A \og_\A \vert 0_\A \rangle \langle 0_\A \vert \oog_\A g'_\A 
\nonumber \\ 
    && \qquad\qquad +
    \beta^* \alpha \:
	  g_\A \oog_\A \vert 0_\A \rangle \langle 0_\A \vert \og_\A g'_\A 
\nonumber \\ 
    && \qquad\qquad +
  \left. \vphantom{\sum} 
    \vert \beta \vert^2 \:
		g_\A \oog_\A \vert 0_\A \rangle \langle 0_\A \vert \oog_\A g'_\A 
  \right]
\nonumber
. 
\nonumber 
\eea
The normalisation factor is determined by requiring that 
${\rm Tr} \rho_\A = 1$, which is straightforwardly obtained from 
Eq.~\eqref{eq: rho proj 2 states bis}: 
\bea
\N &=& 
\vert G_\A \vert 
\left[ 
  1 
	+ 
	x 
  \sum_{g \in G_\A} 
	  \langle 0_\A \vert g_\A \og_\A \oog_\A \vert 0_\A \rangle 
\right] 
, 
\eea
where we used the fact that $\vert \alpha \vert^2 + \vert \beta \vert^2 = 1$ 
and we defined $x = \alpha^* \beta + \beta^* \alpha$ for brevity. 
Notice that the expectation value in the normalisation factor either 
fixes $g_\A$ uniquely if $\og_\A \oog_\A \otimes \openone_\B \in G$, or it 
vanishes identically otherwise. 

Let us consider first the case where an element 
$g^* = \og_\A \oog_\A \otimes \openone_\B$ in $G$ exists (and therefore 
$g^* \in G_\A$). 
We proceed to take the trace over subsystem $\Aii$ 
term by term from Eq.~\eqref{eq: rho proj 2 states bis}, 
leading to the expectation values 
\bea
\langle 0_{\Aii} \vert g'_{\Aii} g_{\Aii} \vert 0_{\Aii} \rangle
\\
\langle 0_{\Aii} \vert \oog_{\Aii} g'_{\Aii} g_{\Aii} 
\og_{\Aii} \vert 0_{\Aii} \rangle
\\
\langle 0_{\Aii} \vert \og_{\Aii} g'_{\Aii} 
g_{\Aii} \oog_{\Aii} \vert 0_{\Aii} \rangle 
\\
\langle 0_{\Aii} \vert g'_{\Aii} g_{\Aii} \vert 0_{\Aii} \rangle 
. 
\eea
The first and last expectation values can always be 
satisfied and they impose $g' = g \tg$, for any $\tg \in G_{\Ai}$. 
On the contrary, the second and third expectation values can be satisfied 
only if an element $h \in G_\A$ exists such that 
$h_{\Aii} = \og_{\Aii} \oog_{\Aii}$. 
Our assumption of the existence of $g^* \in G_\A$ clearly satisfies this 
condition for $h = g^*$. 
However, we notice that this is a sufficient but 
not necessary assumption, and we shall comment more on this later. 

The existence of such element $g^*$ allows us to change the summation label 
$g' \to g' g g^*$ for any $g \in G_{\A}$, whereby: 
\bea
\langle 0_{\Aii} \vert \oog_{\Aii} g'_{\Aii} g_{\Aii} 
\og_{\Aii} \vert 0_{\Aii} \rangle 
&\to& 
\langle 0_{\Aii} \vert g'_{\Aii} \vert 0_{\Aii} \rangle
\nonumber\\
\oog_{\Ai} g'_{\Ai} 
&\to& 
\og_{\Ai} g_{\Ai} g'_{\Ai} 
\nonumber\\ 
\og_{\Ai} g'_{\Ai} 
&\to& 
\oog_{\Ai} g_{\Ai} g'_{\Ai} 
. 
\nonumber
\eea
The new form of the expectation value in the first of the three expressions 
above straightforwardly imposes that $g' \in G_{\Ai}$, and the 
reduced density matrix becomes 
\bea
\rho_{\Ai} 
&=& 
\frac{1}{\N} 
\sum_{g \in G_\A} 
\sum_{\tg \in G_{\Ai}} 
\label{eq: rho_A1 two states}\\ 
&\times& 
  \left[ \vphantom{\sum} 
    \left( 
		  \vert \alpha \vert^2 
      + 
      \alpha^* \beta 
		\right) 
		g_{\Ai} \og_{\Ai} \vert 0_{\Ai} \rangle 
		\langle 0_{\Ai} \vert \og_{\Ai} g_{\Ai} \tg_{\Ai} 
  \right.
\nonumber \\ 
&& \!\!\!\!
+
\left. \vphantom{\sum} 
    \left( 
  		\beta^* \alpha 
			+
			\vert \beta \vert^2
		\right)
	  g_{\Ai} \oog_{\Ai} \vert 0_{\Ai} \rangle 
		\langle 0_{\Ai} \vert \oog_{\Ai} g_{\Ai} \tg_{\Ai} 
  \right]
. 
\nonumber 
\eea
%
%

In order to obtain the von Neumann entropy of subsystem $\Ai$ we begin 
by computing the second power of $\rho_{\Ai}$, 
\begin{widetext}
\bea
\rho^2_{\Ai} 
&=& 
\frac{1}{\N^2} 
\sum_{g,g' \in G_\A} 
\sum_{\tg,\tg' \in G_{\Ai}} 
\left[ \vphantom{\sum} 
  \left( 
	  \vert \alpha \vert^2 
    + 
    \alpha^* \beta 
	\right)^2 
	g_{\Ai} \og_{\Ai} \vert 0_{\Ai} \rangle 
	\langle 0_{\Ai} \vert 
	  g_{\Ai} \tg_{\Ai} g'_{\Ai} 
	\vert 0_{\Ai} \rangle 
	\langle 0_{\Ai} \vert \og_{\Ai} g'_{\Ai} \tg'_{\Ai} 
\right.
\label{eq: rho_A1 two states squared}\\ 
&& \qquad\qquad\qquad\qquad\:
+
  \left( 
	  \vert \alpha \vert^2 
    + 
    \alpha^* \beta 
	\right) 
  \left( 
	  \beta^* \alpha 
		+
		\vert \beta \vert^2
	\right)
	g_{\Ai} \og_{\Ai} \vert 0_{\Ai} \rangle 
	\langle 0_{\Ai} \vert 
	  \og_{\Ai} g_{\Ai} \tg_{\Ai} g'_{\Ai} \oog_{\Ai} 
	\vert 0_{\Ai} \rangle 
	\langle 0_{\Ai} \vert \oog_{\Ai} g'_{\Ai} \tg'_{\Ai} 
\nonumber \\ 
&& \qquad\qquad\qquad\qquad\:
+
  \left( 
	  \beta^* \alpha 
		+
		\vert \beta \vert^2
	\right)
  \left( 
	  \vert \alpha \vert^2 
    + 
    \alpha^* \beta 
	\right) 
	g_{\Ai} \oog_{\Ai} \vert 0_{\Ai} \rangle 
	\langle 0_{\Ai} \vert 
	  \oog_{\Ai} g_{\Ai} \tg_{\Ai} g'_{\Ai} \og_{\Ai} 
	\vert 0_{\Ai} \rangle 
	\langle 0_{\Ai} \vert \og_{\Ai} g'_{\Ai} \tg'_{\Ai} 
\nonumber \\ 
&& \qquad\qquad\qquad\qquad\:
+
\left. \vphantom{\sum} 
  \left( 
  	\beta^* \alpha 
		+
		\vert \beta \vert^2
	\right)^2
  g_{\Ai} \oog_{\Ai} \vert 0_{\Ai} \rangle 
	\langle 0_{\Ai} \vert 
	  g_{\Ai} \tg_{\Ai} g'_{\Ai} 
	\vert 0_{\Ai} \rangle 
	\langle 0_{\Ai} \vert \oog_{\Ai} g'_{\Ai} \tg'_{\Ai} 
\right]
. 
\nonumber 
\eea
\end{widetext}
The first and last term involve the expectation value 
\bea
\langle 0_{\Ai} \vert 
  g_{\Ai} \tg_{\Ai} g'_{\Ai} 
\vert 0_{\Ai} \rangle 
, 
\eea
which can be evaluated with the 
help of the change of variable $g' \to g \tg g'$. This then imposes 
$g'_{\Ai} = \openone_{\Ai}$. 
The second and third terms involve instead the expectation value 
\bea
\langle 0_{\Ai} \vert 
  \oog_{\Ai} g_{\Ai} \tg_{\Ai} g'_{\Ai} \og_{\Ai} 
\vert 0_{\Ai} \rangle 
, 
\eea
which once again vanishes identically unless an element $h \in G_\A$ exists 
such that $h_{\Ai} = \og_{\Ai} \oog_{\Ai}$. 
Our assumption of the existence of $g^* \in G_\A$ clearly satisfies this 
condition for $h = g^*$ (this is again sufficient but 
not necessary), and we can perform the change of variable 
$g' \to g \tg g^* g'$ whereby the expectation value reduces 
to $\langle 0_{\Ai} \vert g'_{\Ai} \vert 0_{\Ai} \rangle$. 
In all cases, the dependence on $g'$ disappears, and it can thus be trivially 
summed over, resulting in an overall factor of $\vert G_{\Aii} \vert$. 

After these steps, we arrive at the expression, 
\begin{widetext}
\bea
\rho^2_{\Ai} 
&=& 
\frac{\vert G_{\Aii} \vert}{\N^2} 
\sum_{g \in G_\A} 
\sum_{\tg,\tg' \in G_{\Ai}} 
\left[ \vphantom{\sum} 
  \left( 
	  \vert \alpha \vert^2 
    + 
    \alpha^* \beta 
	\right)^2 
	g_{\Ai} \og_{\Ai} \vert 0_{\Ai} \rangle 
	\langle 0_{\Ai} \vert \og_{\Ai} g_{\Ai} \tg_{\Ai} \tg'_{\Ai} 
\right.
\\ 
&& \qquad\qquad\qquad\qquad\:
+
  \left( 
	  \vert \alpha \vert^2 
    + 
    \alpha^* \beta 
	\right) 
  \left( 
	  \beta^* \alpha 
		+
		\vert \beta \vert^2
	\right)
	g_{\Ai} \og_{\Ai} \vert 0_{\Ai} \rangle 
	\langle 0_{\Ai} \vert \og_{\Ai} g_{\Ai} \tg_{\Ai} \tg'_{\Ai} 
\nonumber \\ 
&& \qquad\qquad\qquad\qquad\:
+
  \left( 
	  \beta^* \alpha 
		+
		\vert \beta \vert^2
	\right)
  \left( 
	  \vert \alpha \vert^2 
    + 
    \alpha^* \beta 
	\right) 
	g_{\Ai} \oog_{\Ai} \vert 0_{\Ai} \rangle 
	\langle 0_{\Ai} \vert \oog_{\Ai} g_{\Ai} \tg_{\Ai} \tg'_{\Ai} 
\nonumber \\ 
&& \qquad\qquad\qquad\qquad\:
+
\left. \vphantom{\sum} 
  \left( 
  	\beta^* \alpha 
		+
		\vert \beta \vert^2
	\right)^2
  g_{\Ai} \oog_{\Ai} \vert 0_{\Ai} \rangle 
	\langle 0_{\Ai} \vert \oog_{\Ai} g_{\Ai} \tg_{\Ai} \tg'_{\Ai} 
\right]
, 
\nonumber 
\eea
\end{widetext}
where we used the fact that $g^*_{\Ai} \og_{\Ai} = \oog_{\Ai}$ and 
$g^*_{\Ai} \oog_{\Ai} = \og_{\Ai}$. 
We notice that $\tg$ and $\tg'$ always appear as the product 
$\tg \tg'$. A straightforward redefinition $\tg \to \tg \tg'$ allows to 
remove the dependence on $\tg'$, which thus resums to a factor 
$\vert G_{\Ai} \vert$. 
Moreover, the first two terms and the latter two terms differ from one 
another only by their coefficients in curved brackets, which can thus be 
combined to: 
\bea
\left( 
  \vert \alpha \vert^2 
  + 
  \alpha^* \beta 
\right) 
\left( 1 + x \right)
\\
\left( 
  \beta^* \alpha 
	+
	\vert \beta \vert^2
\right) 
\left( 1 + x \right)
, 
\eea
respectively, where we used the normalisation condition 
$\vert \alpha \vert^2 + \vert \beta \vert^2 = 1$ and 
the shorthand notation introduced earlier, 
$x = \alpha^* \beta + \beta^* \alpha$. We finally recognise that 
\bea
\rho^2_{\Ai} 
&=& 
\frac{\vert G_{\Ai} \vert \vert G_{\Aii} \vert}{\N^2} 
\sum_{g \in G_\A} 
\sum_{\tg \in G_{\Ai}} 
\left( 1 + x \right) 
\\ 
&\times& 
  \left[ \vphantom{\sum} 
    \left( 
		  \vert \alpha \vert^2 
      + 
      \alpha^* \beta 
		\right) 
		g_{\Ai} \og_{\Ai} \vert 0_{\Ai} \rangle 
		\langle 0_{\Ai} \vert \og_{\Ai} g_{\Ai} \tg_{\Ai} 
  \right.
\nonumber \\ 
&& +
\left. \vphantom{\sum} 
    \left( 
  		\beta^* \alpha 
			+
			\vert \beta \vert^2
		\right)
	  g_{\Ai} \oog_{\Ai} \vert 0_{\Ai} \rangle 
		\langle 0_{\Ai} \vert \oog_{\Ai} g_{\Ai} \tg_{\Ai} 
  \right]
\nonumber 
\\ 
&=& 
\frac{\vert G_{\Ai} \vert \vert G_{\Aii} \vert}{\N} 
\left( 1 + x \right) 
\; 
\rho_{\Ai}
\\ 
&=& 
\frac{\vert G_{\Ai} \vert \vert G_{\Aii} \vert}{\vert G_\A \vert} 
\; 
\rho_{\Ai}
, 
\eea
which is the same as Eq.~\eqref{eq: rho_A1^2 propto rho_A1} in the main text. 

To summarise, provided that the states in the superposition satisfy the 
condition $\og_\A \oog_\A \otimes \openone_\B \in G$, the result obtained in 
the main text holds: the von Neumann entropy of $\Ai$ after subsystem 
$\B$ has been projected onto the superposition is a direct measure of the 
topological entropy of the system, devoid of any area law or other terms 
scaling with the size of the partitions. 
(We refrain here from verifying that the same is true for the 
entanglement negativity as well.) 

What happens if the superposition \textit{does not} satisfy this condition? 
We need to start again from Eq.~\eqref{eq: rho proj 2 states bis}, 
where we now have $\N = \vert G_{\A} \vert$. 
In this case, the calculations involve significantly longer expressions and 
we do not report them here in full. Once again, we find that $\rho_{\Ai}$ 
depends on the expectation value 
\bea
\langle 0_{\Aii} \vert 
  \oog_{\Aii} g'_{\Aii} g_{\Aii} \og_{\Aii} 
\vert 0_{\Aii} \rangle
, 
\label{eq: exp 1}
\eea
and $\rho^2_{\Ai}$ further depends on the expectation value 
\bea
\langle 0_{\Ai} \vert 
  \oog_{\Ai} g_{\Ai} \tg_{\Ai} g'_{\Ai} \og_{\Ai} 
\vert 0_{\Ai} \rangle 
, 
\label{eq: exp 2}
\eea
for $g,g' \in G_\A$ and $\tg \in G_{\Ai}$. 
Depending on the nature of the elements $\og$ and $\oog$, one can show that 
all possible combinations are allowed, namely: 
\begin{itemize}
\item[(i)] both expectation values admit a non-vanishing solution; 
\item[(ii)] only one of them does; or 
\item[(iii)] both vanish identically. 
\end{itemize}
(We note that option (i) were both admit a solution does \textit{not} 
contradict the assumption $\og_\A \oog_\A \otimes \openone_\B \notin G$.) 

Let us consider in detail scenario (iii) where both expectation values 
above vanish identically for all choices of $g,g' \in G_\A$ and 
$\tg \in G_{\Ai}$. 
As a result, the terms proportional to $\alpha^* \beta$ and to 
$\beta^* \alpha$ in the reduced density matrix $\rho_{\Ai}$ in 
Eq.~\eqref{eq: rho_A1 two states} are absent: 
\bea
\rho_{\Ai} 
&=& 
\frac{1}{\vert G_\A \vert} 
\sum_{g \in G_\A} 
\sum_{\tg \in G_{\Ai}} 
\label{eq: rho_A1 two states case (iii)}
\\ 
&\times& 
  \left[ \vphantom{\sum} 
	  \vert \alpha \vert^2 
		g_{\Ai} \og_{\Ai} \vert 0_{\Ai} \rangle 
		\langle 0_{\Ai} \vert \og_{\Ai} g_{\Ai} \tg_{\Ai} 
  \right.
\nonumber \\ 
&& \!\!\!\!
+
\left. \vphantom{\sum} 
		\vert \beta \vert^2
	  g_{\Ai} \oog_{\Ai} \vert 0_{\Ai} \rangle 
		\langle 0_{\Ai} \vert \oog_{\Ai} g_{\Ai} \tg_{\Ai} 
  \right]
. 
\nonumber 
\eea
The calculation of its second power results in the same terms 
that were obtained in Eq.~\eqref{eq: rho_A1 two states squared}, 
up to trivial changes in the constant coefficients in round brackets: 
\bea
\rho^2_{\Ai} 
&=& 
\frac{\vert G_{\Ai} \vert \vert G_{\Aii} \vert}{\vert G_\A \vert^2} 
\sum_{g \in G_\A} 
\sum_{\tg \in G_{\Ai}} 
\label{eq: rho^2_A1 two states case (iii)}
\\ 
&\times& 
  \left[ \vphantom{\sum} 
	  \vert \alpha \vert^4 
		g_{\Ai} \og_{\Ai} \vert 0_{\Ai} \rangle 
		\langle 0_{\Ai} \vert \og_{\Ai} g_{\Ai} \tg_{\Ai} 
  \right.
\nonumber \\ 
&& \!\!\!\!
+
\left. \vphantom{\sum} 
		\vert \beta \vert^4
	  g_{\Ai} \oog_{\Ai} \vert 0_{\Ai} \rangle 
		\langle 0_{\Ai} \vert \oog_{\Ai} g_{\Ai} \tg_{\Ai} 
  \right]
. 
\nonumber 
\eea

This result can be straightforwardly iterated to give 
\bea
\rho^n_{\Ai} 
&=& 
\frac{\left( \vert G_{\Ai} \vert \vert G_{\Aii} \vert \right)^{n-1}}
     {\vert G_\A \vert^n} 
\sum_{g \in G_\A} 
\sum_{\tg \in G_{\Ai}} 
\label{eq: rho^n_A1 two states case (iii)}
\\ 
&\times& 
  \left[ \vphantom{\sum} 
	  \vert \alpha \vert^{2n} 
		g_{\Ai} \og_{\Ai} \vert 0_{\Ai} \rangle 
		\langle 0_{\Ai} \vert \og_{\Ai} g_{\Ai} \tg_{\Ai} 
  \right.
\nonumber \\ 
&& \!\!\!\!
+
\left. \vphantom{\sum} 
		\vert \beta \vert^{2n}
	  g_{\Ai} \oog_{\Ai} \vert 0_{\Ai} \rangle 
		\langle 0_{\Ai} \vert \oog_{\Ai} g_{\Ai} \tg_{\Ai} 
  \right]
, 
\nonumber 
\eea
and thus 
\bea
{\rm Tr}\left( \rho^n_{\Ai} \right) 
= 
\left( 
  \frac{\vert G_{\Ai} \vert \vert G_{\Aii} \vert}{\vert G_\A \vert} 
\right)^{n-1}
\left[ 
  \vert \alpha \vert^{2n} 
	+ 
	\vert \beta \vert^{2n}
\right]
, 
\nonumber
\eea
from which we obtain the von Neumann entropy: 
\bea
S_{\rm vN}^{(\Ai)} 
&=& 
- \lim_{n \to 1} \partial_n 
\left[ {\rm Tr}\left( \rho^n_{\Ai} \right) \right]
\label{eq: S_vN superp}
\\ 
&=& 
\ln \frac{\vert G_\A \vert}{\vert G_{\Ai} \vert \vert G_{\Aii} \vert} 
- 
\vert \alpha \vert^2 \ln \vert \alpha \vert^2 
- 
\vert \beta \vert^2 \ln \vert \beta \vert^2 
. 
\nonumber
\eea
In addition to the usual topological contribution, in this case a new term 
appears that reminds of the classical entropy of mixing between the two 
states in the superposition (recall that 
$\vert \alpha \vert^2 + \vert \beta \vert^2 = 1$). 

Similarly, although the derivation is not reported here, one can compute 
the entanglement engativity in this case. After a few lines of algebra, 
one arrives at the expressions 
\bea
{\rm Tr} \left[ \left( \rho^{T_2}_{\A} \right)^{2n} \right] 
&=& 
\left( 
  \frac{\vert G_{\Ai} \vert \vert G_{\Aii} \vert}{\vert G_\A \vert} 
\right)^{2n-2}
\nonumber \\ 
&\times&
\left[ 
  \vert \alpha \vert^{2n} 
	+ 
	\vert \beta \vert^{2n}
\right]^2
\\ 
{\rm Tr} \left[ \left( \rho^{T_2}_{\A} \right)^{2n+1} \right] 
&=& 
\left( 
  \frac{\vert G_{\Ai} \vert \vert G_{\Aii} \vert}{\vert G_\A \vert} 
\right)^{2n}
\nonumber \\ 
&\times&
\left[ 
  \vert \alpha \vert^{2(2n+1)} 
	+ 
	\vert \beta \vert^{2(2n+1)}
\right]
. 
\eea
We see once again that the even and odd series lead to different analytic 
continuations. On the one hand, the odd series for $(2n+1) \to 1$ 
(i.e., $n \to 0$) 
tends to $1$, as expected for ${\rm Tr} \left( \rho^{T_2}_{\A} \right)$. 
On the other hand, the even series for $2n \to 1$ 
(i.e., $n \to 1/2$) tends to 
\bea
\| \rho^{T_2}_{\A} \|_1 
&=& 
\frac{\vert G_\A \vert}{\vert G_{\Ai} \vert \vert G_{\Aii} \vert}
\left( 
  \vert \alpha \vert
	+ 
	\vert \beta \vert
\right)^2 
\eea
and we obtain a finite entanglement negativity 
\bea
\E &\equiv& \ln \| \rho^{T_2}_{\A} \|_1 
\nonumber \\ 
&=& 
\ln \frac{\vert G_\A \vert}{\vert G_{\Ai} \vert \vert G_{\Aii} \vert} 
+ 
2 \ln 
\left( 
  \vert \alpha \vert
	+ 
	\vert \beta \vert
\right)
. 
\eea

Both the von Neumann entropy and the entanglement negativity carry an 
additional contribution determined by the coefficients in the superposition, 
on top of the expected topological contribution. 
It is interesting to notice that, although similar in features (e.g., both 
peak at $\ln 2$ when $\vert \alpha \vert = \vert \beta \vert = 1/\sqrt{2}$), 
the two additional contributions are in fact not identical. 

Regarding the two remaninig cases, 
we find through similar calculations 
that the result in the main text holds for (ii), irrespective of which of 
the two expectation values vanishes and which does not. 
Case (i) appears to require significantly more elaborate calculations that 
were not carried out in this work, although we expect that the result 
in the main text holds also in this case. 

In conclusion, we can understand the results in this appendix as follows. 
The existence of a non-trivial subgroup $G_\A \subset G$ (i.e., 
$\vert G_\A \vert > 1$) tells us that the projection of $\B$ onto a 
given state does \emph{not} fully determine the state of the system. 
Indeed, the elements of $G_\A$ label the states in the GS superposition 
that contribute to the projected density matrix. 

Now, let us compare the projection of $\B$ onto two different states 
represented by $\og$ and $\oog$, and let us focus on the case at hand where 
we are interested in measuring the surviving entanglement between $\Ai$ and 
$\Aii$. 
So long as the remaining freedom in the system (i.e., the summation over 
elements of $G_\A$) allows to interchange $\og$ with $\oog$ on at least one 
of the two components of $\A$, then the two projected measures of 
entanglement are in fact one and the same. In mathematical terms, this 
translates into at least one of the following conditions being satisfied, 
\bea
\exists h \in G_{\A} \; &:& \; h_{\Aii} \og_{\Aii} = \oog_{\Aii}
\\
\exists h' \in G_{\A} \; &:& \; h'_{\Ai} \og_{\Ai} = \oog_{\Ai}
, 
\eea
which can in turn be expressed as the expectation values encountered above, 
Eq.~\eqref{eq: exp 1} and Eq.~\eqref{eq: exp 2}: 
\bea
\langle 0_{\Aii} \vert 
  \oog_{\Aii} h_{\Aii} \og_{\Aii} 
\vert 0_{\Aii} \rangle
\\ 
\langle 0_{\Ai} \vert 
  \oog_{\Ai} h'_{\Ai} \og_{\Ai} 
\vert 0_{\Ai} \rangle 
. 
\eea
for $h = g' g$ and $h' = g \tg g'$ ($g,g' \in G_\A$ and 
$\tg \in G_{\Ai} \subset G_\A$). 

Whenever at least one of the two expectation values does not vanish 
(i.e., cases (i)-(ii) above), we are essentially free to interchange 
$\og$ and $\oog$ in the entanglement calculation and one expects the result 
to be the same as in the main text, where only one state was considered. 

On the contrary, when both expectation values vanish identically, the two 
states $\og$ and $\oog$ give inherently different contributions and we find 
that, in addition to the expected topological entropy, the von Neumann 
entropy and entanglement negativity pick up a contribution dependent on the 
chosen quantum superposition of $\og$ and $\oog$. 
This happens, for instance, when $\og = \openone$ and $\oog$ is given by 
the product of two star operators, one acting simultaneously on $\Ai$ and 
$\Bi$, and the other acting simultaneously on $\Aii$ and $\Bii$. 
%
%


\begin{thebibliography}{99}

\bibitem{Kitaev2006}
A. Y. Kitaev and J. Preskill, 
\prl~\textbf{96}, 110404 (2006). 

\bibitem{Levin2006}
M. Levin and X.-G. Wen, 
\prl~\textbf{96}, 110405 (2006).

\bibitem{Zyczkowski1998}
K. Zyczkowski, P. Horodecki,A. Sanpera, and M. Lewenstein, 
\pra~\textbf{58}, 883 (1998); 
see also 
K. Zyczkowski, 
\pra~\textbf{60}, 3496 (1999). 

\bibitem{Lee2000}
J. Lee, M.S. Kim, Y.J. Park, and S. Lee, 
J. Mod. Opt.~\textbf{47}, 2151 (2000). 

\bibitem{Eisert2001}
J. Eisert, PhD thesis, University of Potsdam (2001). 

\bibitem{Plenio2005}
M. B. Plenio, 
\prl~\textbf{95}, 090503 (2005). 
Albeit published at a later date than Ref.~\onlinecite{Lee2000}, it is 
the author's understanding that the work carried out by Eisert and Plenio, 
as well as that by Vidal and Werner, was in fact contemporary 
(and independent) of Lee \textit{et al}. 
For a historical account see Ref.~\onlinecite{Calabrese2013}. 

\bibitem{Vidal2002}
G. Vidal and R. F. Werner, 
\pra~\textbf{65}, 032314 (2002). 

\bibitem{Lee2013}
Y. A. Lee and G. Vidal, 
\pra~\textbf{88}, 042318 (2013). 

\bibitem{Castelnovo2013}
C. Castelnovo, 
\pra~\textbf{88}, 042319 (2013). 

\bibitem{Stephan2012}
J.-M. St\'{e}phan, G. Misguich and V. Pasquier, 
J. Stat. Mech.~\textbf{P02003} (2012). 

\bibitem{Kitaev2003}
A. Y. Kitaev, 
Ann. Phys. (N.Y.) \textbf{303}, 2 (2003).

\bibitem{Hamma2005}
A. Hamma, R. Ionicioiu, and P. Zanardi, 
\pra~\textbf{71}, 022315 (2005).

\bibitem{Castelnovo2008}
C. Castelnovo and C. Chamon, 
\prb~\textbf{77}, 054433 (2008). 

\bibitem{Castelnovo2007}
C. Castelnovo and C. Chamon, 
\prb~\textbf{76}, 174416 (2007). 

\bibitem{Calabrese2012}
P. Calabrese, J. Cardy, E. Tonni, 
\prl~\textbf{109}, 130502 (2012). 

\bibitem{Calabrese2013}
P. Calabrese, J. Cardy, E. Tonni, 
J. Stat. Mech.~\textbf{P02008} (2013); 
P. Calabrese, L. Tagliacozzo, and E. Tonni, 
J. Stat. Mech.~\textbf{P05002} (2013). 

\bibitem{Alba2013}
V. Alba, 
J. Stat. Mech.~\textbf{P05013} (2013). 

\bibitem{eight-vertex_refs}
B. Sutherland, 
J. Math. Phys.~\textbf{11}, 3183 (1970); 
C. Fan and F. Y. Wu, 
\prb~\textbf{2}, 723 (1970). 

\bibitem{Castelnovo2007finT}
C. Castelnovo and C. Chamon, 
\prb~\textbf{76}, 184442 (2007). 

\bibitem{Castelnovo2008finT}
C. Castelnovo and C. Chamon, 
\prb~\textbf{78}, 155120 (2008). 

\bibitem{Lee2013b}
E. K.-H. Lee, R. Schaffer, S. Bhattacharjee, and Y. B. Kim, 
\prb~\textbf{89}, 045117 (2014); 
S.-B. Lee, E. K.-H. Lee, A. Paramekanti, Y. B. Kim, 
arXiv:1309.7050 (2013). 

\bibitem{Kimchi2013}
I. Kimchi, J. G. Analytis, A. Vishwanath, 
arXiv:1309.1171 (2013). 

\bibitem{Nasu2013}
J. Nasu, T. Kaji, K. Matsuura, M. Udagawa, Y. Motome, 
arXiv:1309.3068 (2013). 

\bibitem{Fradkin2006}
E. Fradkin and J. E. Moore, 
\prl~\textbf{97}, 050404 (2006); 
E. Fradkin, 
J. Phys. A: Math. Theor.~\textbf{42}, 504011 (2009). 

\bibitem{Oshikawa2010}
M. Oshikawa, 
arXiv:1007.3739 (2010). 

\end{thebibliography}
\end{document}